\documentclass[a4paper,11pt]{article}
\usepackage{epsfig}

\def\uparrow{\kern0.1em\delimiter"3222378 \kern0.1em}    
\def\downarrow{\kern0.1em\delimiter"3223379 \kern0.1em}  
\def\Uparrow{\kern0.05em\delimiter"322A37E \kern0.05em}
\def\Downarrow{\kern0.05em\delimiter"322B37F \kern0.05em}

\def\underuparrow{\mathrel{\mathop{\uparrow}\limits_{-}}}

\def\underdownarrow{\mathrel{\mathop{\downarrow}\limits_{-}}}

\def\myuparrow{\mathrel{\mathop{\uparrow}\limits_{}}}
\def\myUparrow{\mathrel{\mathop{\Uparrow}\limits_{}}}
\def\mydownarrow{\mathrel{\mathop{\downarrow}\limits_{}}}
\def\myDownarrow{\mathrel{\mathop{\Downarrow}\limits_{}}}

\def\uparrowA{\mathrel{\mathop{\myuparrow}\limits_{\mathrm{A}}}}

\def\DownarrowA{\mathrel{\mathop{\myDownarrow}\limits_{\mathrm{A}}}}

\def\UparrowB{\mathrel{\mathop{\myUparrow}\limits_{\mathrm{B}}}}
\def\downarrowB{\mathrel{\mathop{\mydownarrow}\limits_{\mathrm{B}}}}

\def\underuparrowA{\mathrel{\mathop{\underuparrow}\limits_{\mathrm{A}}}}

\def\underdownarrowB{\mathrel{\mathop{\underdownarrow}\limits_{\mathrm{B}}}}

\def\overline#1{\mathrel{\mathop{#1}\limits^{\cleaders\hbox{$\mkern-2mu\mathord-\mkern-2mu$}\hfill}}}
\def\doubleoverline#1{\mathrel{\mathop{#1}\limits^{\cleaders\hbox{$\mkern-2mu\mathord=\mkern-2mu$}\hfill}}}

\title{Some prospects for ensemble solid-state NMR quantum computers}
\author{\large{A.\ A.\ Kokin, K.\ A.\ Valiev}}
\date{}

\begin{document}

\maketitle

\thanks{Institute of Physics and Technology of RAS, 34, Nakhimovskii pr., 117218 Moscow, Russia}

\begin{abstract}
As an ensemble scheme of solid-state NMR quantum computers the
extension of Kane's many-qubits silicon scheme based on the array of
$^{31}\mathrm{P}$ donor atoms are spaced lengthwise of the strip gates is
considered. The possible planar topology of such ensemble quantum
computer is suggested. The estimation of the output NMR signal was
performed and it was shown that for the number $N \geq 10^{5}$ of ensemble
elements involving $L \sim 10^{3}$ qubits each, the standard NMR methods are
usable.\par
As main mechanisms of decoherence for low temperature ($< 0.1 K$),
the adiabatic processes of random modulation of qubit resonance
frequency determined by secular part of nuclear spin hyperfine
interaction with electron magnetic moment of basic atom and dipole-dipole interaction with nuclear moments of neighboring impurity atoms
was considered, It was made estimations of allowed concentrations of
magnetic impurities and of spin temperature whereby the required
decoherence suppression is obtained. Semiclassical decoherence model
of two qubit entangled states is also presented.\par
As another variant of the solid-state ensemble quantum computer,
the gateless architecture of cellular-automaton with
antiferromagnetically ordered electron spins is also discussed here.\par
\end{abstract}


\section*{Introduction}

Atomic nuclei with spin quantum number $I = 1/2$ are the {\it natural
candidates}\/ for qubits in quantum computers. The early approach to NMR
quantum computers was suggested in 1997 \cite{1,2} and then confirmed in
experiments \cite{3,4}. In this approach several diamagnetic organic
liquids whose individual molecules, having a number of interacted non-equivalent nuclear spins-qubits with $I = 1/2$ and being nearly 
independent on one another where used. They act in parallel as an 
ensemble of {\it almost independent}\/ quantum molecules-microcomputers. In so 
doing the nuclear spins of an individual molecule are described by 
mixed state density matrix of {\it reduced quantum ensemble.}
{\it Initialization}\/ of the nuclear spin states in this case means the 
transformation of mixed state into so called, effective or {\it pseudo-pure}\/ state \cite{1,2,4,5}.\par
The access to individual qubits in a liquid sample is replaced by
simultaneous access to related qubits in all molecules of a bulk
ensemble. Computers of this type are called {\it bulk-ensemble}\/ quantum
computers. The liquid-based quantum computer can operate at {\it room temperature}.
For control and measurements of qubit states the standard NMR technique is used.\par
The principle one-coil scheme of experiment is shown in Fig.\ 1. 
The sample is placed in $\mathbf{t}\mathrm{he}$ constant external magnetic field $\mathbf{B}$ and in 
the alternating (say, linearly polarized) field $\mathbf{b}(t)$, produced by RF 
voltage $V_{\omega}(t)$:
\begin{eqnarray}
\mathbf{B}(t) = \mathbf{B}\mathbf+ \mathbf{b}(t) = B \mathbf{k}\mathbf+ 2b \cos (\omega t + \varphi ) \mathbf{i}\mathbf,\label{1}
\end{eqnarray}
where $\mathbf{i}$ and $\mathbf{k}$ are unit vectors along the axes x and z.\par
\par
\begin{center}
\epsfbox{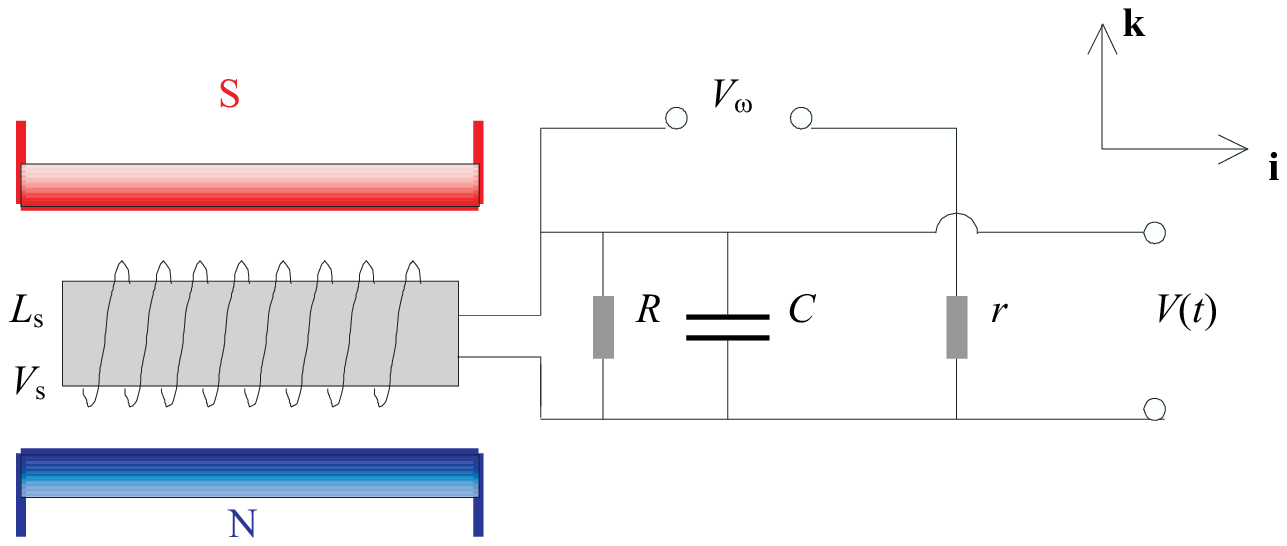}
\nobreak\par\nobreak
Fig.\ 1. The principle one-coil scheme of NMR measurement.\par
\end{center}
\par
Let the sample represent an ensemble of $N$
molecules--microcomputers with $L$ qubits each at temperature
$T = 300\,\mathrm{K}$, in the external magnetic fields $B = 1-10\,\mathrm{T}$. The
resonance nuclear spin frequency is $\omega_{\mathrm{A}}/2\pi \sim \gamma_{\mathrm{I}}B/2\pi < 150\,\mathrm{MHz}$, $\gamma_{\mathrm{I}}$ is
gyromagnetic ratio of nuclear spin ($\gamma_{\mathrm{I}} \sim \gamma_{\mathrm{N}} = 95.8\,\mathrm{radMHz/T}$),
$\hbar \omega_{\mathrm{A}}/kT < 10^{-5}$.\par
The output oscillating voltage $V(t)$ is
\begin{eqnarray}
V(t) = QK d\Phi(t)/dt = \mu_{0} QK A dM_{\mathrm{x}}(t)/dt,\label{2}
\end{eqnarray}
where $\Phi (t) = \int_{A} \mu_{0} M_{\mathrm{x}}(t) dy dz$
is magnetic flux produced by resonant spins
in the coil ($\mu_{0} = 4\pi \cdot 10^{-1} \textrm{T}^{2}\mathrm{cm}^{3}/\mathrm{J})$, $L_{\mathrm{s}} = \mu_{0}(KA)^{2}/V_{\mathrm{s}}$ is solenoid
inductance of the resonance counter, $V_{\mathrm{s}}$ is volume of the solenoid, $K$
is the number and $A$ is area of coil turns, $Q = R/(\omega_{\mathrm{A}}L_{\mathrm{s}}) > 10^{2}$ is the
{\it quality factor}\/ of resonance counter for parallel connected resistance
$R$ (Fig.\ 1). For resonance condition $\omega = \omega_{\mathrm{A}} = (L_{\mathrm{s}}C)^{-1/2}$.\par
The maximum nuclear spin read-out magnetization $M_{\mathrm{xma}\mathrm{x}}$ (the liquid 
sample is considered here to be a continuous medium and to have volume 
$V \sim V_{\mathrm{s}})$ at {\it optimum resonance condition}\/ is defined by the amplitude of 
RF field $b = 1/(\gamma_{\mathrm{I}} \sqrt{T_{\perp \mathrm{I}}T_{\parallel\mathrm{I}}})$ \cite{6}\/ (see also (20) below):
\begin{eqnarray}
M_{\mathrm{xma}\mathrm{x}} = M_{\mathrm{zm}} \sqrt{T_{\perp \mathrm{I}}/T_{\parallel\mathrm{I}}}/2 \approx \gamma_{\mathrm{I}}\hbar /2\cdot (N/V_{\mathrm{s}})\cdot \varepsilon(L)/2, \label{3}
\end{eqnarray}
where $M_{\mathrm{zm}}$ is maximum equilibrium nuclear magnetization, $T_{\perp \mathrm{I}}$ and $T_{\parallel\mathrm{I}}$ 
are effective transverse and longitudinal relaxation times, $N$ is 
number of resonant nuclear spins (one in a molecule) in volume $V_{\mathrm{s}}$. 
Parameter $\varepsilon(L)$ is the maximum probability of the full nuclear 
polarization in pseudo-pure state $P_{\mathrm{I}} = 1$ \cite{7}. It may be estimated by 
the difference of equilibrium population between the lowest and the 
highest energy states. For nearly homonuclear $L$ spin system \cite{7}\/ it is:
\begin{eqnarray}
\varepsilon(L) = \frac{\exp(L\hbar \omega_{\mathrm{A}}/2kT) - \exp (-L\hbar \omega_{\mathrm{A}}/2kT)}{(\exp (\hbar \omega_{\mathrm{A}}/2kT) + \exp (-\hbar \omega_{\mathrm{A}}/2kT))^{L}} = \frac{2\mathrm{sinh}(L\hbar \omega_{\mathrm{A}}/2kT)}{2^{L}\mathrm{cosh}^{L}(\hbar \omega_{\mathrm{A}}/2kT)}.\label{4}
\end{eqnarray}
In the high temperature limit $\hbar \omega_{\mathrm{A}}/(kT) \ll 1$ we have
$\varepsilon(L) = L2^{-L}\hbar \omega_{\mathrm{A}}/(kT)$,
that is, the signal amplitude {\it exponentially drops}\/ with the number of qubits, but it {\it does not}\/ drop for $\hbar \omega_{\mathrm{A}}/(kT) \gg 1$ when
$\varepsilon(L) = 1$ (the pure ground nuclear spin quantum state).\par
The maximum NMR signal intensity $S$ is defined by amplitude
\begin{eqnarray}
S = \left|V_{\mathrm{max}}\right| = (\mu_{0}/4)  Q KA (N/V_{\mathrm{s}}) \gamma_{\mathrm{I}}\hbar \omega_{\mathrm{A}} \varepsilon(L) ,\label{5}
\end{eqnarray}
where the product $KA$ can also be expressed as
\begin{eqnarray}
KA = (L_{\mathrm{s}}V_{\mathrm{s}}/\mu_{0})^{1/2} = (R V_{\mathrm{s}}/(\mu_{0}Q \omega_{\mathrm{A}}))^{1/2}.\label{6}
\end{eqnarray}
For the root-mean square noise voltage in the measurement circuit we write
\begin{eqnarray}
V_{\mathrm{N}} = \sqrt{4 kT R \Delta \nu} ,\label{7}
\end{eqnarray}
where the amplifier bandwidth is as a rule $\Delta \nu \sim 1\,\mathrm{Hz}$.\par
So for {\it signal to noise ratio}\/ we obtain
\begin{eqnarray}
(\mathrm{S}/\mathrm{N}) & \equiv & \left|V_{\mathrm{max}}\right|/V_{\mathrm{N}} \cong \frac{1}{8} \sqrt{ \frac{\mu_{0}\hbar Q \hbar \omega_{\mathrm{A}}}{\Delta \nu V_{\mathrm{s}} kT}} \gamma_{\mathrm{I}}N \varepsilon(L) \sim \nonumber
\\
& \sim & 0.2 \sqrt{(Q/V_{\mathrm{s}})\cdot (\hbar \omega_{\mathrm{A}}/kT)} N \varepsilon(L)\cdot 10^{-9}, (\mathrm{here} \; V_{\mathrm{s}} \; \mathrm{in}\; \mathrm{cm}^{3}).\label{8}
\end{eqnarray}
For example, for two qubits molecules ($L = 2$), using, 
$\varepsilon(L) = \hbar \omega_{\mathrm{A}}/(2kT) \sim 10^{-5}$, we can make an estimation\begin{eqnarray}
(\mathrm{S}/\mathrm{N}) \sim (Q/V_{\mathrm{s}})^{1/2} N\cdot 10^{-16}.\label{9}
\end{eqnarray}
Thus, to keep the value $(\mathrm{S}/\mathrm{N}) > 1$, the number of resonant nuclear 
spins for two qubit liquid ensemble at room temperature, $V_{\mathrm{s}} \sim 1\,\mathrm{cm}^{3}$ 
and $Q \sim 10^{3}$ is bound to be $N > 10^{16}$.\par
In the case of {\it paramagnetic liquids}\/ one would expect that the 
number of polarized nuclei may be increased with dynamic polarization 
(say, Overhauser effect). Assuming electron and nuclear gyromagnetic 
ratio $\gamma_{\mathrm{e}}/\gamma_{\mathrm{I}} \sim 10^{3}$ we obtain that in the probability $\varepsilon(L)$ for a 
$L-$qubits single state the value $\hbar \omega_{\mathrm{A}}/(kT)$ in (\ref{8}) should be 
replaced by $10^{3}\hbar \omega_{\mathrm{A}}/(kT)$. Therefore, for the same value $\varepsilon(L)$ and 
number of molecules $N$, the allowed number of qubits $L$ approximately 
will be estimated from
\begin{eqnarray}
L2^{-L} > 10^{-3},\label{10}
\end{eqnarray}
whence it follows that $L < 12$ qubits.\par
An additional increase of read-out NMR signal may be obtained in 
paramagnetic liquids using the ENDOR technique. It is generally 
believed that for the liquid bulk-ensemble quantum computers a
{\it limiting value}\/ is $L < 20-30$ \cite{7}.\par
There are five basic criteria for realization of a {\it large-scale 
NMR quantum computer}, which can outperform all traditional classical 
computers \cite{17}:
\begin{enumerate}
\item For any physical system, which presents large-scale quantum
register, the necessary number of qubits in quantum register must be
$L > 10^{3}$.\par
One such example of this register is solid-state homonuclear
system, in which nuclear spin containing identical atoms are housed at
regular intervals in a natural or an artificial solid-state structure.\par
\item There is a need to provide the conditions for preparation of
initial basic quantum register state. For a many-qubit solid-state
NMR quantum computer the quantum register state initializing can be
obtained by going to {\it extra-low nuclear spin temperature}\/ ($<1\,\mathrm{mK}$ at
fields of order of several tesla).\par
\item The decoherence time of qubit states $T_{\mathrm{d}}$ should be at least up
to $10^{4}$ times longer than the `clock time', that is value of order of
several seconds for NMR quantum computers. {\it The decoherence suppression
is one of the important problems}\/ in realization of a large-scale
quantum computers.\par
\item There is a need to perform during a decoherence time
a set of quantum logic operations determined by a
logic unitary transformation. This set should contain certain set of
the one-qubit and two-qubits operations are shielded from random
errors. The electromagnetic pulses that control the quantum operation
should be performed with an accuracy of better than $10^{-4}$--$10^{-5}$.\par
\item There is a need to provide {\it accurate and sensitive read-out
measurements of the qubit states}. This is another of the
important and hard problems.
\end{enumerate}
The design of solid-state NMR quantum computers was proposed by
B. Kane in \cite{8,9}. It was suggested to use a semiconductor MOS
structure on a $^{28}\mathrm{Si}$ spinless substrate, in a near-surface layer whose
stable phosphorus isotopes $^{31}\mathrm{P}$, acting as donors, are implanted in the
form of a regular chain. These donors have a nuclear spin $I = 1/2$ and
substitute for silicon atoms at the lattice sites, producing shallow
impurity states. The number of donors or the qubit number $L$ in such a
quasi-one-dimensional {\it artificial `molecule'}\/ may be arbitrary large. It
is suggested an {\it individual}\/ nuclear spin--qubits electrical control
and measurement of qubit states through the use of special gate
structures. The experimental implementation of Kane's scheme is undertaken
now in Australian Centre for Quantum Computer Technology \cite{10,11}.\par
However, there are four essential difficulties in implementing
this quantum computer:\par
\begin{enumerate}
\item First of all, signal from the spin of an individual atom is
very small and {\it high sensitive single-spin measurements}\/ are required.\par
\item For initialization of nuclear spin states it is required to
use {\it very low nuclear spin temperature}\/ ($\sim \,\mathrm{mK}$).\par
\item It is required to use regular donors and gates arrangement
with high precision in {\it nanometer scale}.\par
\item It is necessary to suppress the {\it decoherence}\/ of quantum states
defined by {\it fluctuations}\/ of gate voltage.
\end{enumerate}
As an alternative, we proposed the variant of {\it an ensemble
silicon-based quantum computer}\/ \cite{12,13}. One would expect that with the
ensemble approach, where many independent `molecules' of Kane's type 
work simultaneously, the measurements would be greatly simplified. 
Here we will give some further development of this scheme.\par
\par
\section{The silicon structure with regular system of strip gates}
\par
In this case, unlike the structure suggested in \cite{8}, gates $\mathbf{A}$ and
$\mathbf{J}$ form a chain of narrow ($l_{\mathrm{A}} \sim 10\,\mathrm{nm}$) and long strips along which
donor atoms at $l_{\mathrm{y}}$ distant from each other are placed (Fig.\ 2). Thus, 
they form a {\it regular}\/ structure of the planar silicon topology type.\par
\par
\begin{center}
\epsfbox{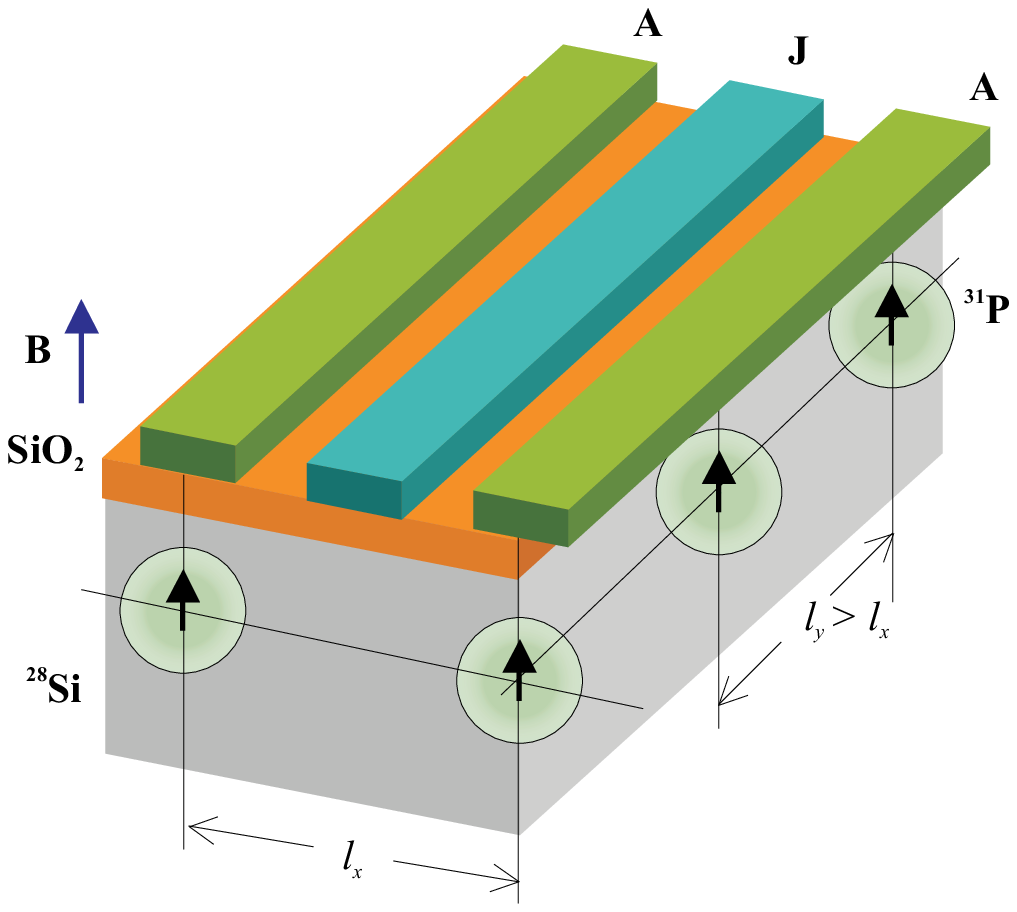}
\nobreak\par\nobreak
Fig.\ 2. The structure of two qubit cells for three ensemble component.\par
\end{center}
\par
The separation between neighboring donor atoms in Si, as in 
Kane's scheme, must be $l_{\mathrm{x}} \leq 20\,\mathrm{nm}$. In this case the interqubit 
interaction is controlled by gates $\mathbf{J}$. The depth of donor $d$ is $\sim 20\,\mathrm{nm}$. 
For $l_{\mathrm{y}} \gg l_{\mathrm{x}}$ the exchange spin interaction between electrons of donor 
atoms disposed along the strip gates ($y-$axis) is negligibly small. 
Hence, such a system breaks down into an ensemble of near-independent 
Kane's artificial `molecule', whose electronic spins at temperature 
$T \leq 0.1\,\mathrm{K}$ are initially fully aligned with the field of several Tesla 
($\gamma_{\mathrm{e}}\hbar B/kT \gg 1$). As in case of liquids, the nuclear spin states of 
individual Kane's chain-`molecule' will be described by density matrix 
of reduced quantum ensemble. Access to individual qubits will be 
replaced by simultaneous access to related qubits in all `molecules'
of ensemble.\par
The linear qubit density in the artificial `molecules' is 
$\sim 50$ qubits on micrometer. For the realization of considered 
structure, as well as of the Kane's scheme, {\it the nanotechnology}\/ with 
resolution of the order of $\sim $ 1 nm is also needed.\par
For {\it the initializing}\/ of all nuclear spin-qubit quantum states 
(fully polarized nuclear spins) there is a need to attain, for the 
time being, nuclear spin temperature $T \leq 10^{-3}\mathrm{K}$. An output signal in 
this system, as in liquids, will be proportional to the number of 
`molecules' or donor atoms $N$ (component number of our ensemble) in the 
chain along axis $y$. In the following the lower value of $N$ will be 
estimated.\par
\par
\par
\section{The states of insulated donor atoms in magnetic fields}
\par
The electron-nuclear spin Hamiltonian for a donor atom $^{31}\mathrm{P}$ has the form
\begin{eqnarray}
H = \gamma_{\mathrm{e}}\hbar \mathbf{BS}\mathbf- \gamma_{\mathrm{I}}\hbar \mathbf{BI}\mathbf+ A \mathbf{IS}\mathbf,\label{11}
\end{eqnarray}
four energy levels of which are given by the well-known Breit-Rabi 
formula. For $I = 1/2$, $S = 1/2$ (the $z-$axis is parallel to $\mathbf{B})$ this 
formula is written as
\begin{eqnarray}
E(F,m_{\mathrm{F}}) = - \frac{A}{4} - \gamma_{\mathrm{I}}\hbar Bm_{\mathrm{F}} - (-1)^{\mathrm{F}} \mathrm{sign}(1+m_{\mathrm{F}}X) \frac{A}{2} \sqrt{1 + 2m_{\mathrm{F}}X + X^{2}} ,\label{12}
\end{eqnarray}
where constant of hyperfine interaction $A/(2\pi \hbar ) = 116\,\mathrm{MHz} \cite{14}$,
$X = (\gamma_{\mathrm{e}} + \gamma_{\mathrm{I}})\hbar B/A \approx \gamma_{\mathrm{e}}\hbar B/A \gg 1$, $F = I \pm 1/2 = 1, 0$, and
$m_{\mathrm{F}} = M + m = \pm 1, 0$, if $F = 1$ or $m_{\mathrm{F}} = 0$, if $F = 0$ (Here $M = \pm 1/2$
and $m = \pm 1/2$ are $z-$projections of electron and nuclear spins
accordingly). The energy level scheme is shown in Fig.\ 3. For the
energy of the ground spin state, $F = 0$ and $m_{\mathrm{F}} = 0$, hence, we obtain
\begin{eqnarray}
E(0,0) = - A/4 - (A/2) \sqrt{1 + X^{2}}.\label{13}
\end{eqnarray}
For the next, excited energy state, $F = 1$, $m_{\mathrm{F}} = -1$ we have
\begin{eqnarray}
E(1,-1) = A/4 - (\gamma_{\mathrm{e}} - \gamma_{\mathrm{I}}) \hbar B/2.\label{14}
\end{eqnarray}
Thus, the energy difference between the two lower states of the 
nuclear spin (the resonant qubit frequency), that interacts with an 
electron, whose state remains unchanged, is described in simple terms 
$(\gamma_{\mathrm{e}} \gg \gamma_{\mathrm{I}}$ for $X \approx \gamma_{\mathrm{e}}\hbar B/A \gg 1)$:
\begin{eqnarray}
\hbar \omega_{\mathrm{A}}^{+} &=& E(1,-1) - E(0,0) = A/2 + (\gamma_{\mathrm{I}} - \gamma_{\mathrm{e}})\mathrm{h}B/2 + \frac{A}{2} \sqrt{1 + X^{2}} \approx \nonumber
\\
&\approx& \gamma_{\mathrm{I}}\hbar B + \frac{A}{2} - \frac{A^{2}}{4\gamma_{\mathrm{e}}\hbar B} ,\nonumber
\\
\hbar \omega_{\mathrm{A}}^{-} &=& E(1,1) - E(1,0) \approx - \gamma_{\mathrm{I}}\hbar B + \frac{A}{2} + \frac{A^{2}}{4\gamma_{\mathrm{e}}\hbar B}.\label{15}
\end{eqnarray}
For $^{31}\mathrm{P}$ donor atoms $\gamma_{\mathrm{e}}/\gamma_{\mathrm{I}} = 1.62\cdot 10^{3}$, $\gamma_{\mathrm{e}} = 176.08\,\mathrm{radGHz/T}$,
$\gamma_{\mathrm{I}} = 1.13\gamma_{\mathrm{N}} = 108\,\mathrm{radMHz/T}$. In magnetic field $B = 1\,\mathrm{T}$: $\omega_{\mathrm{A}}^{+}/2\pi = 75\,\mathrm{MHz}$,
$\omega_{\mathrm{A}}^{-}/2\pi = 41\,\mathrm{MHz}$.\par
\par

\begin{center}
\epsfbox{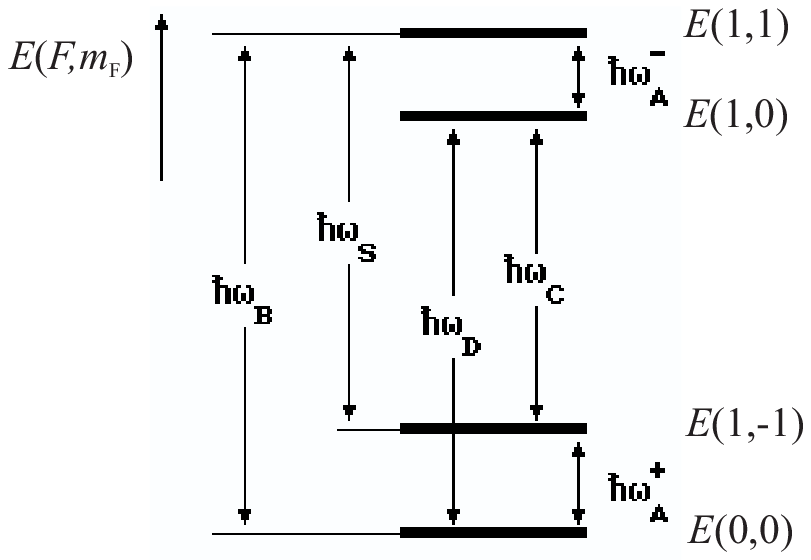}
\nobreak\par\nobreak
Fig.\ 3. Energy levels of an individual donor atom in magnetic field.\par
\end{center}

\par
The frequencies $\omega_{\mathrm{S}}$, $\omega_{\mathrm{B}}$, $\omega_{\mathrm{C}}$, $\omega_{\mathrm{D}}$ are in microwave, $\omega_{\mathrm{A}}^{\pm}$ -- in the RF
ranges of frequencies. The transitions with frequencies $\omega_{\mathrm{S}}$ in the
first approximation are {\it forbidden}.\par
The states $\left|F,m_{\mathrm{F}}\right> $ in $M, m$ basis are 
\begin{eqnarray}
\left|1,1\right> & = & \left|1/2,1/2\right> ,\nonumber
\\
\left|1,-1\right> & = & \left|-1/2,-1/2\right> ,\nonumber
\\
\left|1,0\right> & = & (1-\alpha )^{1/2} \left|1/2,-1/2\right> + \alpha^{1/2}\left|-1/2,1/2\right> ,\nonumber
\\
\left|0,0\right> & = & (1-\alpha )^{1/2} \left|-1/2,1/2\right> - \alpha^{1/2}\left|1/2,-1/2\right> ,\nonumber
\\
\alpha & = & \frac{1}{2} \left(1 - \frac{X}{\sqrt{1 + X^{2}}} \right) \approx 1/(4X^{2}) \ll 1.\label{16}
\end{eqnarray}
The diagonal matrix elements of nuclear magnetization $M_{\mathrm{z}}$ per one 
donor atom for two lower energy states will be determined by
\begin{eqnarray}
\left<0,0\right|M_{\mathrm{z}}\left|0,0\right> & = & \left<0,0\right|I_{\mathrm{z}}\left|0,0\right> \gamma_{\mathrm{I}}\hbar = \frac{X}{\sqrt{1 + X^{2}}} \gamma_{\mathrm{I}}\hbar /2 ,\nonumber
\\
\left<1,-1\right|M_{\mathrm{z}}\left|1,-1\right> & = &  \left<1,-1\right|I_{\mathrm{z}}\left|1,-1\right> \gamma_{\mathrm{I}}\hbar = - \gamma_{\mathrm{I}}\hbar /2.\label{17}
\end{eqnarray}
The probabilities of the $L - $qubit lowest and highest energy 
fully filling states for the same electron spin state $M = - 1/2$, which 
correspond, as noted above, to the maximum probability of the nuclear 
polarization in pseudo-pure state, are:
\begin{eqnarray}
p^{L}(1,-1) & = & \frac{\exp (-L\hbar \omega_{\mathrm{A}}^{+}/2kT)}{(\exp (\hbar \omega_{\mathrm{A}}^{+}/2kT) + \exp (-\hbar \omega_{\mathrm{A}}^{+}/2kT))^{L}} ,\nonumber
\\
p^{L}(0,0) & = & \frac{\exp (L\hbar \omega_{\mathrm{A}}^{+}/2kT)}{(\exp (\hbar \omega_{\mathrm{A}}^{+}/2kT) + \exp (-\hbar \omega_{\mathrm{A}}^{+}/2kT))^{L}}.\label{18}
\end{eqnarray}
The possible maximum nuclear magnetization $M_{\mathrm{zm}}$ (the populations
of states $\left|1,1\right>$ and $\left|1.0\right>$ is negligible for $\omega_{\mathrm{S}}, \omega_{\mathrm{B}}, \omega_{\mathrm{C}} \gg \omega_{\mathrm{A}}^{\pm})$ is
\begin{eqnarray}
M_{\mathrm{zm}} & = & \gamma_{\mathrm{I}}\hbar /2\cdot (N/V_{\mathrm{c}}) \left( \frac{X}{\sqrt{1 + X^{2}}} \frac{\exp (L\hbar \omega_{\mathrm{A}}^{+}/2kT)}{(\exp (\hbar \omega_{\mathrm{A}}^{+}/2kT) + \exp (-\hbar \omega_{\mathrm{A}}^{+}/2kT))^{L}} - \right. \nonumber
\\
& - & \left. \frac{\exp (-L\hbar \omega_{\mathrm{A}}^{+}/2kT)}{(\exp (\hbar \omega_{\mathrm{A}}^{+}/2kT) + \exp (-\hbar \omega_{\mathrm{A}}^{+}/2kT))^{L}} \right)
= \gamma_{\mathrm{I}}\hbar /2\cdot (N/V_{\mathrm{c}}) \varepsilon(L).\label{19}
\end{eqnarray}
For $L\hbar \omega_{\mathrm{A}}^{+}/2kT \ll 1$ and $X \gg 1$ we obtain (compare with (\ref{3}))
\begin{eqnarray}
M_{\mathrm{zm}} \approx \gamma_{\mathrm{I}}\hbar /2\cdot (N/V_{\mathrm{c}})\cdot 2^{-L}L(\hbar \omega_{\mathrm{A}}^{+}/kT).\label{20}
\end{eqnarray}
But for very low temperatures ($\hbar \omega_{\mathrm{A}}^{+}/2kT \gg 1$) we have the {\it full nuclear polarization}
$M_{\mathrm{zm}} \approx \gamma_{\mathrm{I}}\hbar /2\cdot (N/V_{\mathrm{c}})$ and $\varepsilon(L) = 1$.\par
\par
\section{The gain effect for NMR signal}
\par
Transitions between two lower states are induced by a RF magnetic 
field, applied at a frequency resonant $\omega_{\mathrm{A}}^{+}$. The Rabi resonance 
frequency $\Omega $, which is defined by matrix elements of spin interaction 
Hamiltonian with the external RF field $\mathbf{b}(t)$
\begin{eqnarray}
H_{\mathrm{rf}}(t) = (\gamma_{\mathrm{e}}S_{\mathrm{x}} - \gamma_{\mathrm{I}}I_{\mathrm{x}})\hbar b_{\mathrm{x}}(t) , b_{\mathrm{x}}(t) = 2b\cos (\omega_{\mathrm{A}}^{+}t)\label{21}
\end{eqnarray}can be found from
\begin{eqnarray}                                     
\Omega = \gamma_{\mathrm{I}}b_{\mathrm{eff}}(X) = 2 \left|\left<0,0\right|H_{\mathrm{rf}}(0)\left|1,-1\right>\right|/\hbar.\label{22}
\end{eqnarray}
For the amplitude of effective RF field, acting on nuclear spin, 
$b_{\mathrm{eff}}(X)$ we obtain
\begin{eqnarray}
b_{\mathrm{eff}}(X) = b \left( \alpha^{1/2}(\gamma_{\mathrm{e}}/\gamma_{\mathrm{I}}) + (1 - \alpha )^{1/2} \right) ,\label{23}
\end{eqnarray}
where $b$ is the amplitude of circularly polarized field component.\par
The Rabi frequency has the maximum value for $X = 0$ ($\alpha = 1/2$) and 
monotonically reduces to value for the insulated nuclear spin ($\alpha \Rightarrow 0$), 
$\gamma_{\mathrm{I}}b_{\mathrm{eff}}(X \gg 1) = \gamma_{\mathrm{I}}b$. From the rate of quantum operation standpoint 
it is desirable to operate in relatively weak fields \cite{9}, at which 
$\gamma_{\mathrm{e}}/\gamma_{\mathrm{I}} \gg X \approx \gamma_{\mathrm{e}}\hbar B/A \gg 1$
or $3.5\,\mathrm{T} > B \gg 3.9\cdot 10^{-3}\,\mathrm{T}$.\par
In this case from (\ref{23}) we will obtain
\begin{eqnarray}
b_{\mathrm{eff}} = (1 + \eta )b \gg b,\label{24}
\end{eqnarray}
where $\eta = A/(2\gamma_{\mathrm{I}}\hbar B) \gg 1$ is the {\it gain factor}. Under these conditions RF
field operates through the transverse component of electronic
polarization. For magnetic fields $B = 1\,\mathrm{T}$ we have the value
$b_{\mathrm{eff}} = 4.4\cdot b$, and for $B = 0.01\,\mathrm{T}$ we have the value $b_{\mathrm{eff}} = 338\cdot b$. The
gain effect involves an increase of NMR signal and Rabi frequency.
This effect was indicated previously by K. Valiev in \cite{15}.\par
In the pulse technique this effect makes it possible {\it to decrease
the length of pulse}\/ and along with it the times of logic operation
performing. Moreover, the computer operations, owing to this effect,
can be performed at lower RF fields. At last, it permits to {\it reduce the
RF field influence}\/ on the operation of neighboring semiconductor
devices.\par
To describe the nuclear dynamics for the two low-lying level
systems being discussed ($X \gg 1$), we can write the following Bloch-type equation with only two effective relaxation times:
\begin{eqnarray}
\frac{d\mathbf{M}}{dt} = \gamma_{\mathrm{I}} [\mathbf{M}\mathbf\times \mathbf{B}_{\mathrm{eff}}] - \frac{M_{\mathrm{x}}\mathbf{i}\mathbf+ M_{\mathrm{y}}\mathbf{j}}{T_{\perp \mathrm{I}}} - \frac{(M_{\mathrm{z}} - M_{\mathrm{zm}})\mathbf{k}}{T_{\parallel\mathrm{I}}} ,\label{25}
\end{eqnarray}
where $\mathbf{i}$, $\mathbf{j}$, $\mathbf{k}$ are orthogonal unit vectors (Fig.\ 1), $M_{\mathrm{zm}}$ is defined as (\ref{19}),
\begin{eqnarray}
\mathbf{B}_{\mathrm{eff}} = (\omega_{\mathrm{A}}/\gamma_{\mathrm{I}}) \mathbf{k}\mathbf+ 2b_{\mathrm{eff}}\cos (\omega t) \mathbf{i}\mathbf.\label{26}
\end{eqnarray}
It follows from it that the value of maximum nuclear read-out
magnetization in NMR signal has here for
$b_{\mathrm{eff}}(X) = 1/(\gamma_{\mathrm{I}}\sqrt{ T_{\perp \mathrm{I}}T_{\parallel\mathrm{I}}})$,
that is again for
$M_{\mathrm{xma}\mathrm{x}} = M_{\mathrm{zm}}\sqrt{ T_{\perp \mathrm{I}}/T_{\parallel\mathrm{I}}}/2$.
Hence, the read-out {\it NMR signal can not be increased}\/ through the gain effect over its maximum value, that
corresponds to
$M_{\mathrm{zm}}\sqrt{ T_{\perp \mathrm{I}}/T_{\parallel\mathrm{I}}}/2$.
\par

\section{The signal to noise ratio for an ensemble silicon quantum computer}
\par
For the realization of an ensemble silicon quantum register we 
propose a variant of planar scheme \cite{16}, that, as an example, contains 
$n\cdot p$ in parallel acting identical blocks, each has $N_{0}$ in parallel 
connected $L - $qubit Kane's linear `molecules'. This scheme is 
schematically depicted in Fig.\ 4.\par
Let the sample be the silicon ($^{28}\mathrm{Si}$) plate of thickness 0.1 cm. 
For the full number of computers-`molecules' in ensemble $N = p\cdot N_{0}\cdot n$, 
the volume of sample and also of solenoid is $V_{\mathrm{s}} \approx \delta \cdot l_{\mathrm{x}}\cdot l_{\mathrm{y}}\cdot L\cdot N$ (the 
filling factor is assumed for simplicity to be one).\par
\par
\par
\begin{center}
\epsfbox{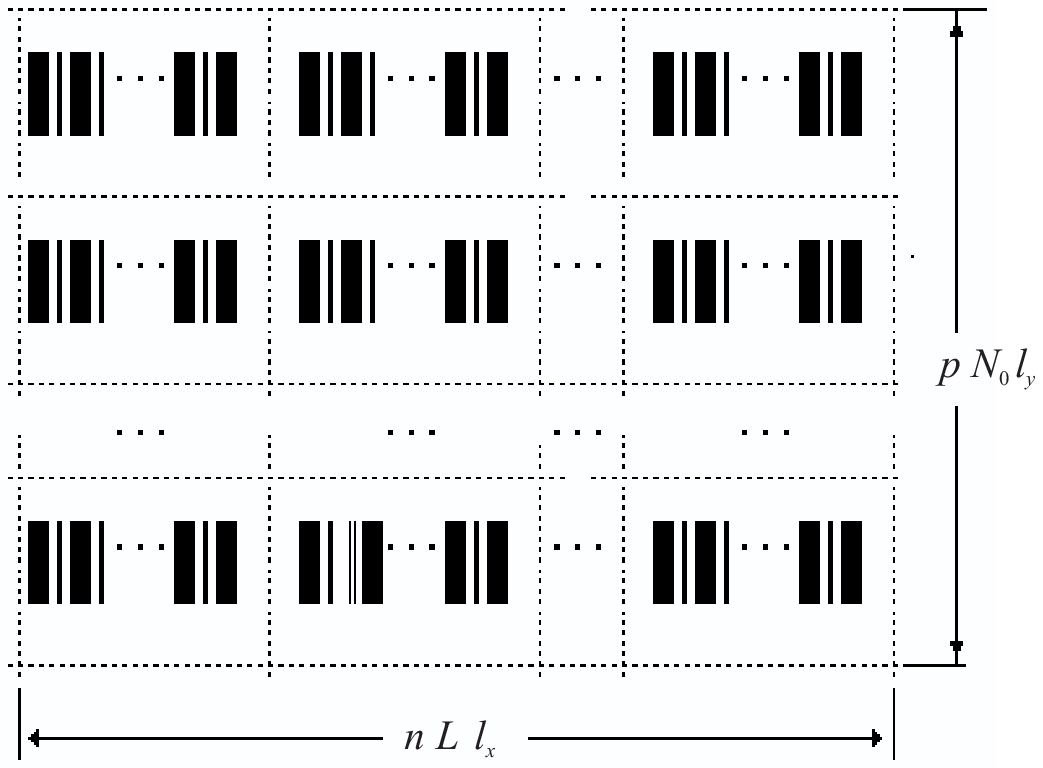}
\nobreak\par\nobreak
Fig.\ 4. The scheme of the proposed planar silicon topology with $p\cdot n$ in
parallel connected blocks of the ensemble $L - $qubit quantum computers
(the connections are not shown here). The broad and narrow lines
denote the $\mathbf{A}$ and $\mathbf{J}$ gates.\par
\end{center}

\par
The read-out signal from such {\it ensemble in parallel acting}\/ chains,
as distinct from liquid prototype, for full nuclear polarization or,
what is the same, for nuclear spin temperatures $T_{\mathrm{I}} \leq  10^{-3}\,\mathrm{K}$ has
instead of the small factor in intensity of the NMR signal of type
$\varepsilon(L) = 2^{-L}L\cdot \hbar \omega_{\mathrm{A}}/(kT)$ the factor $\varepsilon(L) = 1$. The NMR signal from our 
sample within a non-essential factor is the same as from macroscopic 
sample (see Appendix $\mathbf{A}.\mathbf{1})$. Therefore, with the expressions (\ref{4}) and 
(\ref{8}), $\hbar \omega_{\mathrm{A}}/(kT_{\mathrm{I}}) < 1$ ($T_{\mathrm{I}} < 1\,\mathrm{mK}$) and $\varepsilon = 1$ we will obtain as an 
estimation for maximum signal to noise ratio
\begin{eqnarray}
(\mathrm{S}/\mathrm{N}) \approx \sqrt{Q\hbar \omega_{\mathrm{A}}/(kTV_{\mathrm{s}})}\cdot N\cdot 10^{-9} \approx \sqrt{QN/(\delta l_{\mathrm{x}}l_{\mathrm{y}}L)}\cdot 10^{-10}.\label{27}
\end{eqnarray}
It is believed that for low temperatures $Q \sim 10^{6}$. The effective 
volume of one `molecule' for $l_{\mathrm{x}} = 20\,\mathrm{nm}$, $l_{\mathrm{y}} = 50\,\mathrm{nm}$, $L = 10^{3}$, 
$V_{\mathrm{s}} = \delta l_{\mathrm{x}}l_{\mathrm{y}}L = 10^{-9}\,\mathrm{cm}^{3}$ we receive that the read-out signal in our 
scheme may be {\it available for standard NMR technique}, if the number of 
`molecules' in ensemble is of about $N \geq 10^{5}$. So high-sensitive devices 
for measurement of individual spin-states are not needed.\par
To estimate the values $n$, $p$ let us consider the square plate with 
50$N_{0}p = 20\cdot 10^{3} n$ and $N_{0} = 100$. As a result, we receive $n \approx 16$ and 
$p \approx 63.\mathrm{The}$ area of the structure without passive regions is $\sim $ 
315 $\times 315$ $\mu \mathrm{m}^{2}$. This size is sufficiently small for sample to be housed 
in the split between the magnet poles of a standard NMR spectrometer. 
Real plate may have considerably more area and correspondingly more 
number of `molecules' $N$.\par
For implementation of two-qubit logic operation it is required 
the controlled by gates $\mathbf{J}$ interqubit indirect interaction with 
characteristic frequency $\nu_{\mathrm{J}} \sim 100\,\mathrm{kHz} \ll \omega_{\mathrm{A}}/2\pi \sim 100\,\mathrm{MHz}$. To bring 
about fault-tolerant quantum computations on large-scale quantum 
computers the relative error for single logic operation must not be 
more than $\sim 10^{-5}$\cite{17}. Hence it follows that a {\it resolution}\/ bound of the 
NMR spectrometer must be of the order of $\sim 100\,\mathrm{kHz}\cdot 10^{-5} \sim 1\,\mathrm{Hz}$, that 
is consistent with the usual requirements. It is significant that such 
high precision is needed only for performing the logic quantum 
operation, but it is not needed for read-out measurements.\par
The read-out signal may be more increased by means of an 
electron-nuclear double resonance (ENDOR) methods \cite{18} of observing 
the electron resonance at transition with frequencies $\omega_{\mathrm{B}}$ and $\omega_{\mathrm{C}}$ 
(Fig.\ 3).\par
Consequently, by the use of standard NMR and additional of ENDOR 
techniques {\it the first main difficulty}\/ of Kane's scheme can be overcome.\par
\par
\section{The cooling of nuclear spin system and nuclear state 
initialization by means of dynamic polarization}
\par
The electron and nuclear {\it longitudinal relaxation times}\/ for the 
{\it allowed}\/ transitions in four energy level system of phosphorus doped 
silicon have been extensively investigated experimentally in \cite{18,19}. 
For the allowed transitions with frequency $\omega_{\mathrm{B}}$ and $\omega_{\mathrm{C}}$ (Fig.\ 3) electron
longitudinal relaxation times $\tau_{\parallel\mathrm{B}} \approx \tau_{\parallel\mathrm{C}}$ at low temperatures were found
to be exceedingly long. They are of the order of {\it one hour}\/ at 
$T = 1.25\,\mathrm{K}$, $B \sim 0.3\,\mathrm{T}$, are independent of phosphorus concentration 
below $C \sim 10^{16}\,\mathrm{cm}^{-3}$ (mean distance between phosphorus atoms is of the 
order of 45 nm) and are approximately inversely proportional to the 
lattice temperature $T$. The nuclear longitudinal relaxation time $T_{\parallel}$ 
(the frequency $\omega_{\mathrm{A}}^{+})$ were found to be equal to 10 hours.\par
The relaxation time for transition with frequency $\omega_{\mathrm{D}}$, which 
involves a simultaneous electron-nuclear spin flip-flop, at 
$T = 1.25\,\mathrm{K}$, $C \sim 10^{16}\,\mathrm{cm}^{-3}$ and $B \sim 0.3\,\mathrm{T}$ was
$\tau_{\parallel\mathrm{D}} \sim 30\,\mathrm{hours} \gg \tau_{\parallel\mathrm{B}}, \tau_{\parallel\mathrm{C}}$.\par
The {\it extremely long relaxation times}\/ of the electron and nuclear 
spins imply that the required initializing of nuclear quantum states 
(full nuclear {\it nonequilibrium polarizations}) can be attained by deep 
cooling of short duration of {\it only nuclear spin system}\/ to $T_{\mathrm{I}} \leq 1\,\mathrm{mK}$
without deep cooling of the lattice. There is the possibility to reach 
it at the {\it indirect cooling}\/ of nuclear spin system by means of {\it dynamic 
nuclear spin polarization techniques}\cite{19}.\par
One such method of dynamic nuclear spin polarization for donor 
atoms is based on the saturation by {\it the microwave pumping of the 
forbidden transition}\/ (frequency $\omega_{\mathrm{S}}$ in Fig.\ 3), that is designated as
the Abragam's {\it solid state effect}\cite{6,19}.\par
Let us consider this effect as applied to the ensemble of $^{31}\mathrm{P}$ 
atoms. The polarization of electrons $P_{\mathrm{S}} = 2\left< S_{\mathrm{z}}\right>$ and of nuclei 
$P_{\mathrm{I}} = 2\left< I_{\mathrm{z}}\right>$ may be for the sake of simplicity expressed as
\begin{eqnarray}
P_{\mathrm{S}} & = & p(1,1) + p(1,0) - p(1,-1) - p(0,0) ,\nonumber
\\
P_{\mathrm{I}} & = & p(1,1) + p(0,0) - p(1,0) - p(1,-1) ,\label{28}
\end{eqnarray}
where $p(F,m_{\mathrm{F}})$ are the populations of states $\left|F,m_{\mathrm{F}}\right>$ (Fig.\ 3). They also
fulfill the requirement
\begin{eqnarray}
p(1,1) + p(1,0) + p(1,-1) + p(0,0) = 1.\label{29}
\end{eqnarray}
The rate equations for the populations are
(it is assumed, that the relaxation rates for transitions at frequencies
$\omega_{\mathrm{A}}^{\pm}$ are equal to $T_{\parallel\mathrm{A}}$):
\begin{eqnarray}
dp(0,0)/dt & = &  (p(1,1) - p(0,0)r_{\mathrm{B}})/\tau_{\parallel\mathrm{B}}+ (p(1,0) - p(0,0)r_{\mathrm{D}})/\tau_{\parallel\mathrm{D}} +\nonumber
\\
&& + (p(1,-1) - p(0,0)r_{\mathrm{A}}^{+})/T_{\parallel\mathrm{A}} ,\nonumber
\\
dp(1,-1)/dt & = & (p(1,0) - p(1,-1)r_{\mathrm{C}})/\tau_{\parallel\mathrm{C}} + (p(1,1) - p(1,-1))\cdot W_{\mathrm{e}}+\nonumber
\\
&& +  (p(0,0)r_{\mathrm{A}}^{+} - p(1,-1))/T_{\parallel\mathrm{A}} ,\nonumber
\\
dp(1,0)/dt & = &  (p(1,-1)r_{\mathrm{C}}- p(1,0))/\tau_{\parallel\mathrm{C}} + p(0,0)r_{\mathrm{D}}-p (1,0))/\tau_{\parallel\mathrm{D}} +\nonumber
\\
&& + (p(1,1) - p(1,0)r_{\mathrm{A}}^{-})/T_{\parallel\mathrm{A}} ,\nonumber
\\
dp(1,1)/dt & =&  (p(0,0)r_{\mathrm{B}} - p(1,1))/\tau_{\parallel\mathrm{B}} + (p(1,-1) - p(1,1))\cdot W_{\mathrm{e}} +\nonumber
\\
&& + (p(1,0)r_{\mathrm{A}}^{-} - p(1,1))/T_{\parallel\mathrm{A}} ,\label{30}
\end{eqnarray}
where parameters $r_{\mathrm{B},\mathrm{C},\mathrm{D},\mathrm{A}} =\,\exp (-\hbar \omega_{\mathrm{B},\mathrm{C},\mathrm{D},\mathrm{A}}/kT)$ are ratio of rates for 
an up and down thermal transitions. For values $\hbar \omega_{\mathrm{B},\mathrm{C},\mathrm{D}}/kT \gg 1$, 
$\hbar \omega^{\pm}_{\mathrm{A}}/kT \ll 1$ ($T \leq 0.1\,\mathrm{K}$) there are the thermal electron $P_{\mathrm{S}0} \approx - 1$ and
nuclear $P_{\mathrm{I}0} = \hbar \omega_{\mathrm{A}}^{+}/kT \ll 1$ polarizations.\par
Let us assume next that the rate of {\it induced forbidden electron
transitions} $\left|1,1\right> \Rightarrow \left|1,-1\right>$ at frequency $\omega_{\mathrm{S}}$, that is $W_{\mathrm{S}}$ and electron
longitudinal relaxation times satisfy the conditions:
\begin{eqnarray}
W_{\mathrm{S}}^{-1} < \tau_{\parallel \mathrm{B}} \approx \tau_{\parallel \mathrm{C}} \ll T_{\parallel \mathrm{A}}, \tau_{\parallel \mathrm{D}}, \tau_{\parallel \mathrm{S}} ,\label{31}
\end{eqnarray}
where $\tau_{\parallel \mathrm{D}}$, $\tau_{\parallel \mathrm{S}}$ are the longitudinal relaxation times of electron spins
for forbidden transition. Hereafter we shall write
\begin{eqnarray}
dp(0,0)/dt & = & p(1,1)/\tau_{\parallel\mathrm{B}} + (p(1,-1) - p(0,0))/T_{\parallel\mathrm{A}} ,\nonumber
\\
dp(1,-1)/dt & = & p(1,0)/\tau_{\parallel\mathrm{B}} + ((p(1,1) - p(1,-1))\cdot W_{\mathrm{e}} + (p(0,0) - p(1,-1))/T_{\parallel\mathrm{A}} ,\nonumber
\\
dp(1,0)/dt & =&  - p(1,0)/\tau_{\parallel\mathrm{B}} + (p(1,1) - p(1,0))/T_{\parallel\mathrm{A}}, \nonumber
\\
dp(1,1)/dt & =&  - p(1,1)/\tau_{\parallel\mathrm{B}} + (p(1,-1) - p(1,1))\cdot W_{\mathrm{e}} + (p(1,0) - p(1,1))/T_{\parallel\mathrm{A}}\label{32}
\end{eqnarray}
With equations (\ref{28}),(\ref{29}),(\ref{31}) we can obtain the rate equations 
for $P_{\mathrm{S}}$ and $P_{\mathrm{I}}$:
\begin{eqnarray}
dP_{\mathrm{S}}/dt & = & - (P_{\mathrm{S}} + P_{\mathrm{I}})\cdot W_{\mathrm{e}} - (P_{\mathrm{S}} + 1)/\tau_{\parallel\mathrm{B}} ,\nonumber
\\
dP_{\mathrm{I}}/dt & = & - (P_{\mathrm{S}} + P_{\mathrm{I}})\cdot W_{\mathrm{e}} - P_{\mathrm{I}}/T_{\parallel\mathrm{A}}.\label{33}
\end{eqnarray}
The steady-state saturation condition ($W_{\mathrm{e}} \gg 1/T_{\parallel\mathrm{A}}$) of the 
transition $\left|1,1\right> \Rightarrow \left|1,-1\right> $ gives rise to the equalization of the 
populations $p(1,1) = p(1,-1)$ and to the {\it full nuclear spin polarization}\/ respectively
\begin{eqnarray}
P_{\mathrm{I}} = - P_{\mathrm{S}} = p(0,0) = 1.\label{34}
\end{eqnarray}
It is obvious that this state is equivalent to the state with 
nuclear spin temperature $T_{\mathrm{I}} < \hbar \omega_{\mathrm{A}}/k \sim 10^{-3}\,\mathrm{K}$.\par
Let us estimate finally the needed microwave power for 
saturation. The rate of external microwave field that induces
forbidden electron transitions $W_{\mathrm{S}}$ differs from the rate of allowed
transitions with a flip of only one nuclear spin $W$ by small factor
that is proportional to $(B_{\mathrm{S}}/B)^{2}$ \cite{6}, where for isotrope hyperfine
interaction $B_{\mathrm{S}}$ is the field due to the dipole-dipole interaction
between nuclear and electron spins of $^{31}\mathrm{P}$ atom. Let us write
\begin{eqnarray}
W_{\mathrm{S}} \sim (\gamma_{\mathrm{e}}b_{\mathrm{mw}})^{2}\cdot \tau^{*}_{\perp \mathrm{S}}/2 ,\label{35}
\end{eqnarray}
where $(\overline{\Delta \omega^{2}_{\mathrm{S}}})^{1/2} \approx 2\tau^{*-1}_{\perp \mathrm{S}}$ is the {\it nonhomogeneous broadened}\/ resonance line
width for the saturated electron transition, $\tau^{*}_{\perp \mathrm{S}}$, is effective
transverse relaxation time of electron spins, $b_{\mathrm{mw}}$ is the amplitude of
{\it microwave field}.
\par
As a result the saturation condition takes the form
\begin{eqnarray}
W_{\mathrm{S}} > 1/\tau_{\parallel \mathrm{S}},
T_{\parallel \mathrm{A}}
& \mathrm{or} &
W > 1/\tau_{\parallel \mathrm{B}},
\; \; \;
(\gamma_{\mathrm{e}}b_{\mathrm{mw}})^{2}\cdot \tau^{*}_{\perp \mathrm{S}}\tau_{\parallel \mathrm{B}} > 1 ,\label{36}
\end{eqnarray}that is the same form as for allowed transition. By using expression
for quality factor $Q_{\mathrm{c}}$ of microwave cavity
\begin{eqnarray}
Q_{\mathrm{c}} \approx \omega_{\mathrm{S}}b^{2}_{\mathrm{mw}}\cdot V_{\mathrm{r}}/(2\mu_{0}P) ,\label{37}
\end{eqnarray}
where $V_{\mathrm{r}}$ is volume of microwave resonator, $P$ is dissipated power we 
will obtain the following saturation condition
\begin{eqnarray}
W_{\mathrm{e}} \gg 1/T_{\parallel\mathrm{S}}
& \mathrm{or} &
(\gamma_{\mathrm{e}}b_{\mathrm{mw}})^{2} T^{*}_{\perp \mathrm{S}}T_{\parallel\mathrm{S}} \gg 1.\label{36a}
\end{eqnarray}
The dissipated power in cavity for $(\gamma_{\mathrm{e}}b_{\mathrm{mw}})^{2}\cdot \tau^{*}_{\perp \mathrm{S}}\tau_{\parallel \mathrm{B}} = 1$ is
determined by
\begin{eqnarray}
P > \omega_{\mathrm{S}}V_{\mathrm{r}}W(\overline{\Delta\omega^{2}_{\mathrm{S}}})^{1/2}/(2\mu_{0}Q_{\mathrm{c}}\gamma_{\mathrm{e}}^{2}).\label{37a}
\end{eqnarray}
For example, taking $W \sim 1/\tau_{\parallel \mathrm{B}} \sim 10^{3}\,\mathrm{s}^{-1}$, $\omega_{\mathrm{S}} \sim 100\,\mathrm{radGHz}$,
$V_{\mathrm{r}} \sim 1\,\mathrm{cm}^{3}$, $Q_{\mathrm{c}} \sim 1000$ and $(\overline{\Delta\omega^{2}_{\mathrm{S}}})^{1/2} \sim 10^{8}\,\mathrm{s}^{-1}\cite{20}$ as a rough estimate
we obtain $P > 1\,\mathrm{mW}$. Notice that this power is applied only during the
saturation process over the time $\geq W_{\mathrm{e}}^{-1} \sim 1\,\mathrm{ms}$ in the act of qubit
state initialization.\par
Hence, the initialization of nuclear states may be obtained by
using ENDOR technique at the lattice temperature of the order of $0.1\,\mathrm{K}$
and by this means that {\it the second difficulty}\/ of Kane's scheme can be 
overcome.\par
Notice here that there is also another possibility of ensemble 
NMR implementation, which does {\it not have the gate system}. The 
selectivity of nuclear resonance frequencies for individual qubit in 
the ensemble of Kane's chains can be achieved, rather than using the 
$\mathbf{A}-$gate voltage, with the applying of the external magnetic field 
gradients along axis $x$. For neighboring qubits separated by $\sim $ 20 nm
it is required $dB_{\mathrm{z}}/dx \sim 1\,\mathrm{T}/\mathrm{cm}$ (that is feasible now), which produce a
resonance frequency difference $\sim 100\,\mathrm{Hz}$.\par
There was proposed previously an all silicon NMR quantum computer
where qubits are $^{29}\mathrm{Si}$ nuclear spins arranged as chains in a $^{28}\mathrm{Si}$
matrix \cite{20}\/ and with natural crystals of calcium hydroxyapatite,
involving one-dimensional hydrogen chains \cite{21}, in both cases nuclear
resonance frequencies are separated by a magnetic field gradient.
However, in this case too large field gradient, of the order of $1\,\mathrm{T}/\mu\mathrm{m}$ is required.\par
\par
\section{The nuclear spin states decoherence due to hyperfine interaction of 
nuclear and electron spins}
\par
The relaxation of nonequilibrium state of the nuclear spin system
represented by the product of independent (nonentangled) one-qubit
states, owing to the interaction with isotropic environment, shows two 
processes. One is a slow establishment of equilibrium state associated 
with dissipation of energy. For it the diagonal elements of density 
matrix decay with characteristic longitudinal (spin-lattice) 
relaxation time $T_{\parallel}$. The decay of non-diagonal matrix elements called 
decoherence of quantum states is characterized by a decoherence time 
$T_{\mathrm{d}}$ or transverse (spin-spin) relaxation time $T_{\perp}$. The longitudinal 
relaxation times $T_{\parallel}$ in the case of nuclear spin of $^{31}\mathrm{P}$ atoms as qubits 
is defined mainly by thermal modulation of qubit resonance frequency  
accompanied by spin flips. It is usual that for solids $T_{\perp} \ll T_{\parallel}$.\par
The {\it internal adiabatic decoherence}\/ mechanisms due to a random 
modulation of qubit resonance frequency, produced by local fluctuating 
magnetic fields without spin flips. These fields are determined by 
secular parts of interactions of nuclear spins with electron spin of 
the basic phosphorus atoms, with impurity paramagnetic atoms and also 
with nuclear spins of impurity atoms. We have named this 
mechanism as {\it internal}. It seem to be the leading one.\par
The modulation of nuclear spin resonance frequency $\Delta \omega (t)$, which 
is determined by the secular part of hyperfine interaction, may be
written as
\begin{eqnarray}
\Delta \omega (t) = A(t)S_{\mathrm{z}}(t) - A_{0}\left<S_{\mathrm{z}}\right> \approx A_{0}(S_{\mathrm{z}}(t) -\left<S_{\mathrm{z}}\right>) - \Delta A(t)\left<S_{\mathrm{z}}\right>,\label{38}
\end{eqnarray}
where $A(t) = A_{0} + \Delta A(t)$, $\Delta A(t)$ is the modulation of hyperfine
interaction constant, $A_{0} = 725\,\mathrm{radMHz}$. The influence of gate voltage
noise on this frequency modulation was studied in \cite{8,9,23}\/ and it is 
not treated here ({\it external}\/ decoherence process).\par
Another (internal) modulation mechanism of $A(t)$ is the 
interaction of donor atoms with acoustic phonons. It is our belief 
that for very low temperature this mechanism is not essential \cite{24}.\par
Let us consider now the first term in (\ref{38}). We shall follow the 
semiclassical model of {\it adiabatic decoherence of one-qubit state}\/ (Appendix {\bf A.2}).
The correlation function of frequency modulation
$\Delta \omega_{\mathrm{S}}(t) = A_{0}(S_{\mathrm{z}}(t) -\left<S_{\mathrm{z}}\right>)$ is determined by the fluctuations of electron
spin polarization and depends on electron resonance frequency $\omega_{\mathrm{S}}$,
longitudinal $\tau_{1}$ (hours) and transverse $\tau_{2}$ relaxation times. In 
adiabatic case $\omega_{\mathrm{S}} = \gamma_{\mathrm{S}}B >1/\tau_{2} \gg 1/\tau_{1}$ and we will obtain:
\begin{eqnarray}
\left<\Delta_{\mathrm{S}}\omega (t)\Delta \omega_{\mathrm{S}}(0)\right> = \left<\Delta \omega_{\mathrm{S}}^{2}\right>\cdot \exp (-t/\tau_{1}),\label{39}
\end{eqnarray}
where
\begin{eqnarray}
\left<\Delta \omega_{\mathrm{S}}^{2}\right> = A_{0}^{2}\left(\left<S_{\mathrm{z}}^{2}\right> - \left<S_{\mathrm{z}}\right>^{2}\right) = A_{0}^{2}(1 - \tanh^{2}(\gamma_{\mathrm{S}}\hbar B/2kT))/4.\label{40}
\end{eqnarray}
Now, according to (\ref{A.11}), we obtain
\begin{eqnarray}
\Gamma (t) = \left<\Delta \omega_{\mathrm{S}}^{2}\right>\tau_{1}^{2}(t/\tau_{1} - 1 +\,\exp (-t/\tau_{1})).\label{41}
\end{eqnarray}
For $\tau_{1} \approx 10^{4}\,\mathrm{s}$ and $t \sim T_{\mathrm{d}} = 1\,\mathrm{s}$, $1 \ll \left<\Delta \omega_{\mathrm{S}}^{2}\right>\tau_{1}^{2} < (\tau_{1}/T_{\mathrm{d}})^{2}$
we have the non-Markovian random process (slow dampening fluctuations). 
In this case
\begin{eqnarray}
\Gamma (t) = \left<\Delta \omega_{\mathrm{S}}^{2}\right> t^{2}/2\label{42}
\end{eqnarray}
and the effective decoherence time can by estimated from
$T_{\mathrm{d}} \sim \left<\Delta \omega_{\mathrm{S}}^{2}\right>^{-1/2}$.\par
The necessary value of decoherence time for the NMR quantum
computer clock time $\sim 10^{-4}\,\mathrm{s}$ should not exceed several seconds.
Therefore, let us write the requirement for $\gamma_{\mathrm{e}}\hbar B/kT \gg 1$ in the form
\begin{eqnarray}
1/T_{\mathrm{d}}^{2} \approx A_{0}^{2}(1 - \tanh^{2}(\gamma_{\mathrm{e}}\hbar B/2kT))/4 \approx 2A_{0}^{2}\exp (-\gamma_{\mathrm{e}}\hbar B/kT) < 1\,\mathrm{s}^{-2},\label{43}
\end{eqnarray}from which we find that the decoherence suppression will be achieved
only at {\it sufficiently large B}/$T > 30\,\mathrm{T}/\mathrm{K}$. It corresponds to $B = 2\,\mathrm{T}$ for 
lattice temperatures $T < 0.06\,\mathrm{K}$.\par
\par
\section{The adiabatic nuclear spin states decoherence due to interaction 
with nuclear spins of impurity atoms.}
\par
The paramagnetic impurity atoms having magnetic moments play also 
a role of environment for nuclear spins in solid state. However 
decoherence mechanism due to dipole-dipole interaction of their 
magnetic moments with nuclear spins-qubits is suppressed to a large 
extent at $B/T > 30\,\mathrm{T}/\mathrm{K}$ thanks to near-full electron spin polarization 
\cite{24}.\par
Another mechanism of one qubit state decoherence is dipole-dipole 
interaction with not fully polarized nuclear spins $I \neq 0$ of impurity
diamagnetic atoms having concentration $C_{\mathrm{I},\mathrm{imp}}$. Isotope $^{29}\mathrm{Si}$ with 
$\gamma_{\mathrm{I},\mathrm{imp}} = - 53\,\mathrm{radMHz/T}$ is one of such atoms. The random fluctuating 
local field, produced by nuclear spins of impurity atoms has the form
\begin{eqnarray}
\Delta B_{\alpha}(t) = - {\sum_{\mathrm{i},\beta}^{\mathrm{N}}}D_{\alpha ,\beta}(\mathbf{r}_{\mathrm{i}})(I_{\beta ,\mathrm{imp}}(\mathbf{r}_{\mathrm{i}},\mathrm{t}) - \left<I_{\beta ,\mathrm{imp}}(\mathbf{r}_{\mathrm{i}})\right>)\label{44}
\end{eqnarray}
where
\begin{eqnarray}
D_{\alpha ,\beta}(\mathbf{r}_{\mathrm{i}}) = \frac{\mu_{0}}{4\pi} \frac{\gamma_{\mathrm{I}}\gamma_{\mathrm{I},\mathrm{imp}}}{r^{3}_{\mathrm{i}}} \left( \delta_{\alpha \beta} - \frac{3r_{\mathrm{i}\alpha}r_{\mathrm{i}\beta}}{r_{\mathrm{i}}^{2}} \right) ,\label{45}
\end{eqnarray}
$\mathbf{r}_{\mathrm{i}}$ is the distance-vector to $i$-th impurity nuclear spin.\par
In this case correlation function of frequency modulation 
\begin{eqnarray}
 \left<\Delta \omega_{\mathrm{S}}(t)\Delta \omega_{\mathrm{S}}(0)\right> = \gamma_{\mathrm{I}}^{2}\left<B_{\mathrm{z}}(t)B_{\mathrm{z}}(0)\right>
= C_{\mathrm{I},\mathrm{imp}} \int {\sum\limits_{\beta}} D_{\mathrm{z},\beta}(\mathbf{r})(I_{\beta ,\mathrm{imp}}(\mathbf{r},\mathrm{t})I_{\beta ,\mathrm{imp}}(\mathbf{r}) - \left<I_{\beta ,\mathrm{imp}}(\mathbf{r})\right>^{2})d\mathbf{r}\label{46}
\end{eqnarray}
takes the form
\begin{eqnarray}
\left<\Delta_{\mathrm{S}}\omega (t)\Delta \omega_{\mathrm{S}}(0)\right> = \left<\Delta \omega^{2}\right>\exp (-t/T_{\parallel,\mathrm{imp}}) ,\label{47}
\end{eqnarray}
where $T_{\parallel,\mathrm{imp}} \approx 10^{4}\,\mathrm{s}$ is impurity nuclear spin longitudinal relaxation 
time of isotope $^{29}\mathrm{P}$ at low temperature \cite{18}. Taking $T_{\parallel,\mathrm{imp}}$ to be much 
more than $T_{\mathrm{d}} \sim 1\,\mathrm{s}$, for the determination of allowable impurity 
concentration we obtain equation
\begin{eqnarray}
1/T_{\mathrm{d}}^{2} \approx C_{\mathrm{I},\mathrm{imp}}\cdot \frac{(\mu_{0}\gamma_{\mathrm{I}}\gamma_{\mathrm{I},\mathrm{imp}}\hbar )^{2}}{60\pi a^{3}}\cdot \left(1 - \tanh^{2}\left(\left|\gamma_{\mathrm{I},\mathrm{imp}}\right|\hbar B/2kT_{\mathrm{I}}\right)\right),\label{48}
\end{eqnarray}
where $a$ is minimal distance to impurity nuclear spin which for Si is 
of the order of $5\cdot 10^{22}\mathrm{cm}^{-3}$.\par
For $B/T > 30\,\mathrm{T}/\mathrm{K}$ and for spin temperature $T_{\mathrm{I}}$ at which there is 
near-full polarization of nuclear spins
\begin{eqnarray}
\left|\gamma_{\mathrm{I},\mathrm{imp}}\hbar B/kT_{\mathrm{I}}\right| > 1\label{49}
\end{eqnarray}or for $T_{\mathrm{I}} < 0.8\,\mathrm{mK}$, we will obtain that the allowed concentration of 
the isotope $^{29}\mathrm{Si}$ is
\begin{eqnarray}
C_{\mathrm{I},\mathrm{imp}}\% < 4.5\cdot 10^{-2}\%.\label{50}
\end{eqnarray}
This value can be increased due to the further decrease of nuclear 
spin temperature $T_{\mathrm{I}}$. For comparison, natural abundance of isotope $^{29}\mathrm{Si}$ 
in natural silicon is 4.7\%. At present the realized degree of cleaning
$^{28}\mathrm{Si}$ is 99.98\%, which does {\it not fully suit}\/ for our purposes yet.\par
\par
\section{Adiabatic decoherence of entangled two qubit states}
\par
In the processes of input of information and logic operation 
performance some nonentangled initializing states of quantum register 
become entangled. The adiabatic process of transverse relaxation may 
be also the main decoherence mechanism of coherent entangled quantum 
states.\par
As a simple example let us consider here the adiabatic 
decoherence of the pure fully entangled two qubit triplet state EPR-type
$\left|\psi_{\mathrm{EPR}}\right> = \sqrt{1/2} \left(\left|\uparrow\downarrow\right> + \left|\downarrow\uparrow\right>\right)$
with the zeroth projection of the
total spin on z-axis, which has density matrix
\begin{eqnarray}
\rho_{\mathrm{EPR}} = \left|\psi_{\mathrm{EPR}}\right>\left<\psi_{\mathrm{EPR}}\right| = \frac{1}{2}\left(
\begin{tabular}{c c c c }
$0$ & $0$ & $0$ & $0$\\
$0$ & $1$ & $1$ & $0$\\
$0$ & $1$ & $1$ & $0$\\
$0$ & $0$ & $0$ & $0$\\
\end{tabular}
\right).\label{51}
\end{eqnarray}
The action of the environment on qubit states will be described
quasiclassically as correlated random modulation of the qubits
resonance frequencies $\Delta \omega_{1,2}(t)$ and of indirect spin-spin interaction
parameter $\Delta \omega (t) = \Delta I_{\mathrm{I}}(t)/2\mathbf{.}$ The secular part of Hamiltonian for
interaction with the environment is represented by
\begin{eqnarray}
\mathbf{H}(t) &=& - \Delta \omega_{1}(t)(\sigma_{1\mathrm{z}}\otimes \mathbf{1})/2 - \Delta \omega_{2}(t)(\mathbf{1}\otimes \sigma_{2\mathrm{z}})/2 +\nonumber
\\
              & & + \Delta \omega_{I}(t)(\sigma_{1\mathrm{z}}\otimes \sigma_{2\mathrm{z}})/2 ,\label{52}
\end{eqnarray}
where $\sigma_{1\mathrm{z},2\mathrm{z}}$ are Pauli matrix.\par
The density matrix (\ref{51}) under the action of random field in
rotating frame with resonance frequency $\omega_{0}$ is described by expression
\begin{eqnarray}
\rho_{\mathrm{EPR}}(t) = \mathbf{U}(t)^{-1} \rho_{\mathrm{EPR}} \mathbf{U}(t).\label{53}
\end{eqnarray}
In the considered case unitary matrix 4$\times 4$ $\mathbf{U}(t)$ is (
$\varphi_{1,2}(t)=\int_{0}^{t}\Delta\omega_{1,2}dt$,
$\varphi_{\mathrm{I}}(t)=\int_{0}^{t}\Delta\omega_{\mathrm{I}}dt$
):
\begin{eqnarray}
\mathbf{U}(t) &=& (\cos (\varphi_{1}(t)/2)\mathbf{1}\mathbf+i\sin ((\varphi_{1}(t)/2)\sigma_{1\mathrm{z}})\otimes (\cos (\varphi_{2}(t)/2)\mathbf{1}\mathbf+i\sin ((\varphi_{2}(t)/2)\sigma_{2\mathrm{z}}\cdot \nonumber
\\
&& \cdot (\cos \varphi_{\mathrm{I}}(t)(\mathbf{1} \otimes \mathbf{1}) + i\sin \varphi_{\mathrm{I}}(t)(\sigma_{1\mathrm{z}}\otimes \sigma_{2\mathrm{z}})).\label{54}
\end{eqnarray}
For perturbed density matrix we obtain
\begin{eqnarray}
\rho_{\mathrm{EPR}}(t) = \frac{1}{2}\left(
\begin{tabular}{c c c c }
$0$ & $0$ & $0$ & $0$\\
$0$ & $1$ & $\exp (-i(\varphi_{1}(t)-\varphi_{2}(t)))$ & $0$\\
$0$ & $\exp (i(\varphi_{1}(t)-\varphi_{2}(t)))$ & $1$ & $0$\\
$0$ & $0$ & $0$ & $0$\\
\end{tabular}
\right).\label{55}
\end{eqnarray}
We see that the {\it modulation of spin-spin interaction has no effect}\/ on density matrix of triplet EPR-state.\par
Let us assume now that the random phases $\varphi_{1,2}(t)$ have mean value
$\left<\varphi_{1,2}(t)\right> = 0$ and belong to the reduced statistical ensemble, which is
described by {\it two-dimension Gaussian distribution}:
\begin{eqnarray}
w(\varphi_{1}(t),\varphi_{2}(t)) &=& \frac{1}{2\pi \sigma_{1}(t)\sigma_{2}(t)\sqrt{(1-\rho^{2}_{12}(t))}}\cdot \nonumber
\\
&& \cdot \exp \left\{  - \frac{1}{2(1-\rho^{2}_{12}(t))}\cdot \left(\frac{\varphi_{1}^{2}(t)}{\sigma_{1}^{2}(t)} - \frac{2\rho_{12}(t)\varphi_{1}(t)\varphi_{2}(t)}{\sigma_{1}(t)\sigma_{2}(t)} + \frac{\varphi_{2}^{2}(t)}{\sigma_{2}^{2}(t)} \right)
\right\}.\label{56}
\end{eqnarray}
Here 
\begin{eqnarray}
\sigma^{2}_{1,2}(t) = \left<\varphi^{2}_{1,2}(t)\right> = 2\int_{0}^{t}(t-\tau )f_{1,2}(\tau )d\tau \nonumber
\end{eqnarray}are the variances and
\begin{eqnarray}
\rho_{12}(t) = \frac{\left<\varphi_{1}(t)\varphi_{2}(t)\right>}{\sigma_{1}(t)\sigma_{2}(t)} = \frac{2\int_{0}^{t}(t-\tau )f_{12}(\tau )d\tau}{\sigma_{1}(t)\sigma_{2}(t)} ,\label{57}
\end{eqnarray}
where
\begin{eqnarray}
f_{1,2}(\tau ) = \left<\Delta \omega_{1,2}(\tau )\Delta \omega_{1,2}(0)\right>,
& &
f_{12}(\tau ) =\left<\Delta \omega_{1}(\tau )\Delta \omega_{2}(0)\right>.\label{58}
\end{eqnarray}
The normalized mutual correlation function $\rho_{12}(t)$ takes values in
interval
\begin{eqnarray}
0 \leq \rho_{12}(t) \leq 1.\nonumber
\end{eqnarray}
After an averaging (\ref{55}) with (\ref{56}) we have
\begin{eqnarray}
\left<\rho_{\mathrm{EPR}}(t)\right> &=& \frac{1}{2}\left(
\begin{tabular}{c c c c }
$0$ & $0$ & $0$ & $0$\\
$0$ & $1$ & $\exp (-\Gamma (t))$ & $0$\\
$0$ & $\exp (-\Gamma (t))$ & $1$ & $0$\\
$0$ & $0$ & $0$ & $0$\\
\end{tabular}
\right),\label{59}
\end{eqnarray}
where
\begin{eqnarray}
\exp (-\Gamma (t)) & = & \int_{-\infty}^{\infty}d\varphi_{1} \int_{-\infty}^{\infty}d\varphi_{2} w(\varphi_{1},\varphi_{2})\exp (\pm i(\varphi_{1} - \varphi_{2})) =\nonumber
\\
 & = & \exp \{ -(\sigma_{1}^{2}(t) -2\sigma_{1}(t)\sigma_{2}(t)\rho_{12}(t) + \sigma_{2}^{2}(t)\}.\label{60}
\end{eqnarray}
In the absence of random field correlation $\rho_{12}(t)$ = 0 decrement $\Gamma (t)$
is equal to the {\it sum of decrements of two}\/ one qubit states:
\begin{eqnarray}
\Gamma (t) = (\sigma_{1}^{2}(t) + \sigma_{2}^{2}(t))/2 = 2 \Gamma_{1}(t).\label{61}
\end{eqnarray}
In case of maximum correlation $\rho_{12}(t)$ = 1 and $\varphi_{1}(t)$ = $\varphi_{2}(t)$ (the same
mode acts on both qubits) {\it adiabatic decoherence disappears}. Analogous 
properties have the singlet EPR state.\par
We see here, that decoherence of interacted qubits states may 
differ essentially from one qubit decoherence. Under the action of 
fully correlated random fields the coherence of two mentioned 
entangled states is not violate and they may be considered as the 
basis of decoherence-free substrate for logical qubits coding. Clearly 
the pure nonfully entangled states $\left|\psi\right> = (\sqrt{1-\alpha}\left|\uparrow \downarrow\right> + \sqrt{\alpha}\left|\downarrow \uparrow\right>)$ have no
such properties.\par
Adiabatic decoherence of other two qubits fully entangled quantum 
Bell states $\left|\psi\right> = \sqrt{1/2} \left(\left|\uparrow \uparrow \right>\right. \pm \left.\left|\downarrow \downarrow \right>\right)$
under the action of fully correlated random fields with
$\rho_{12}(t)$ = 1 now does not disappear.
Its decrement is now equal to
$\Gamma (t) = (\sigma_{1}(t)$ + $\sigma_{2}(t))^{2}/2  = 2 \sigma^{2}_{1} = 4 \Gamma_{1}(t)$,
that is {\it four times larger}\/ than for one qubit decoherence.\par
\par

\section{An antiferromagnet-based ensemble NMR quantum computer of cellular-automaton type}
\par
For implementation of ensemble silicon quantum computer, 
operating on cellular-automaton principle, it may be usable the 
previously considered ensemble of long chains of donor atoms $^{31}\mathrm{P}$ 
disposed in silicon, but free of the $\mathbf{A}$ and $\mathbf{J}$ gates.\par
If exchange interaction constant (it is here positive) for 
localized electronic spins of $^{31}\mathrm{P}$ along the chain is more than Zeeman 
energy $J(l) \gg \gamma_{\mathrm{e}}\hbar B\tau \sim 6.5\cdot 10^{-23}\,\mathrm{J}$, that corresponds to the distance 
between donors $l_{\mathrm{x}} \sim 20\,\mathrm{nm}$ \cite{9}, that is the electron critical 
temperature (Neel temperature) will be $T_{\mathrm{NS}} \sim 4\,\mathrm{K}$. and the lattice 
temperature is well below the critical temperature for electron 
ordering $T < T_{\mathrm{NS}} \sim J(l)/k$ ($k = 1.38\cdot 10^{-23}\mathrm{J}/\mathrm{K}$ is the Boltzmann 
constant), than the one-dimensional {\it antiferromagnetically ordered 
ground state}\/ of electronic spins can be produced.\par
Due to hyperfine interaction nuclear spins will be oriented 
according to the electronic spin direction in the resultant field and 
can form array with the alternating orientation of nuclear spins. At 
magnetic fields $B < A/\gamma_{\mathrm{I}}\hbar \sim 3.5\,\mathrm{T}$ and at spin temperatures $T \sim 10^{-3}\,\mathrm{K}$ 
the nuclear spins $^{31}\mathrm{P}$ will form a periodic ground state array of 
ABAB$\ldots $type: $\uparrow  \downarrow  \uparrow  \downarrow  \ldots $, where $\uparrow $ marks the ground state of nuclear 
spin in an A-site and $\downarrow $ is the ground state of nuclear spin in a B-site with almost 100\% opposite orientation
($\omega_{\mathrm{A},\mathrm{B}}/kT \geq 1$). That is the
distinct nuclear spins will be {\it in initialized ground state}.\par
Notice that the using of dynamic methods makes possible
the high orientation of nuclear spins also at larger lattice temperatures,
this state will be {\it the long-lived nonequilibrium}\/ nuclear spin state.\par
The nuclear resonant frequencies $\omega_{\mathrm{A},\mathrm{B}}$ of neighboring nuclear 
spins are different for each of the magnetic one-dimensional subarray 
A and B in the chain as they depend on the states of neighboring 
spins. We will take it here in the simple form \cite{13,25,26}:
\begin{eqnarray}
\omega_{\mathrm{A},\mathrm{B}} \equiv \omega (m_{<}+ m_{>}) \approx \left|\gamma_{\mathrm{I}}\hbar B \pm A/2 - I_{\mathrm{n}}\cdot (m_{<} + m_{>})\right|/\hbar ,\label{38a}
\end{eqnarray}
where $I_{\mathrm{n}}$ is constant of two neighboring nuclear indirect spin-spin 
interaction, $m_{<}$ = $\pm 1/2$ and $m_{>}$ = $\pm 1/2$ are the magnetic quantum 
numbers for the left and right nuclear spins. The nonsecular part of 
nuclear-nuclear interaction is neglected here taking into account that 
$\gamma_{\mathrm{I}}\hbar B$, $A/2 \gg I_{\mathrm{n}}$. The difference between nuclear resonant frequencies 
for the distinct neighboring spin orientations is 
$\Delta \omega_{\mathrm{I}}/2\pi \sim I_{\mathrm{n}}/2\pi \hbar \sim 0.5\,\mathrm{MHz}$, whereas the resonant nuclear frequencies 
are $\omega_{\mathrm{A},\mathrm{B}}/2\pi \sim A/4\pi \hbar \sim 120\,\mathrm{MHz}$.\par
For the organization of logic operations let us use the 
addressing to spin states, similar to the scheme put forward in \cite{27}. 
Each nuclear spin in A-site of this scheme has {\it two internal}\/ eigenstates -- ground
$\left|\uparrow\right>$ and excited $\left|\Downarrow\right>$ and in B-site --
$\left|\downarrow\right>$ and $\left|\Uparrow\right>$
accordingly.\par
We take into account that the life time of excited states (the
longitudinal nuclear spin relaxation time $T_{\parallel})$ at low temperatures is
very long. Each {\it logic qubit}\/ of quantum information in this state will
be encoded here, similar to \cite{27}, by the {\it states}\/ of {\it four physical
spin}-{\it qubits}: the logical qubit basis state "0" will be encoded by unit
$\left|\Downarrow\Uparrow\uparrow\downarrow\right>$, while the state "1" will be encoded by
$\left|\uparrow\downarrow\Downarrow\Uparrow\right>$.
It is important here that the resonance frequencies of nuclear spins depend
on neighboring spin states. Both logical states have two excited spin
states and {\it zero projection}\/ of total nuclear spin.\par
Notice that a random inversion of only one spin will result in
degradation of the qubit state. But to form the rough error, for
example, of "0" $\Rightarrow$ "1" type in the coding of stored quantum information
it is essential to invert {\it four}\/ spins {\it simultaneously}. Therefore, it may
be concluded that the considered way of qubit coding ensures {\it a better 
fault-tolerance}\/ with respect to this type of errors.\par
The input and output of the information in the array of ground 
states spins could be performed at the ends of the array, where the 
nuclear spins (say in A-site at the left end) have only one 
neighboring spin and resonant frequency $\omega_{\mathrm{A}}(-1/2)$ ($m_{<} + m_{>} = -1/2$). The
corresponding selective resonance $\pi_{\mathrm{A},-1/2}-$pulse inverts only one 
nuclear spin (in A-site) at the end of array and doesn't influence
other ones. Then the new selective $\pi_{\mathrm{B},0}-$pulse will invert next nuclear
spin (in B-site), which has the opposite orientation of ground and
excited neighbor nuclear spin ($m_{<} + m_{>} = 0$ in A-site) and consequently
the new resonant frequency, distinguished from the frequency of spins
with the neighbor nuclear spin in the same ground states ($m_{<} + m_{>} = 1$)
(See Table).\par
\ \par
\ \par
Table. The $\pi -$pulses for spins in A- and B-sites\par
\nobreak\ \par\nobreak
$
\begin{tabular}{c c c c c c }
Neighbor spin states $ \mathrm{A}-$site & A $\downarrow $ & $\downarrow \,\mathrm{A} \downarrow $ & $\downarrow \,\mathrm{A} \Uparrow $ & $\Uparrow \,\mathrm{A} \downarrow $ & $\Uparrow \,\mathrm{A} \Uparrow $  \\
 & & & & & \\
Resonance frequency & $\nu_{\mathrm{A}}(-1/2)$ & $\nu_{\mathrm{A}}(-1)$ & $\nu_{\mathrm{A}}(0)$ & $\nu_{\mathrm{A}}(0)$ & $\nu_{\mathrm{A}}(\ref{1})$ \\
 & & & & & \\
$\pi -$pulses & $\pi_{\mathrm{A},-1/2}$ & $\pi_{\mathrm{A},-1}$ & $\pi_{\mathrm{A},0}$ & $\pi_{\mathrm{A},0}$ & $\pi_{\mathrm{A},1}$ \\
 & & & & & \\
Neighbor spin states $ \mathrm{B}-$site & B $\downarrow $ & $\uparrow \,\mathrm{B} \uparrow $ & $\uparrow \,\mathrm{B} \Downarrow $ & $\Downarrow \,\mathrm{B} \uparrow $ & $\Downarrow \,\mathrm{B} \Downarrow $  \\
 & & & & & \\
Resonance frequency & $\nu_{\mathrm{B}}(1/2)$ & $\nu_{\mathrm{B}}(\ref{1})$ & $\nu_{\mathrm{B}}(0)$ & $\nu_{\mathrm{B}}(0)$ & $\nu_{\mathrm{B}}(-1)$\\
 & & & & & \\
$\pi -$pulses & $\pi_{\mathrm{B},1/2}$ & $\pi_{\mathrm{B},1}$ & $\pi_{\mathrm{B},0}$ & $\pi_{\mathrm{B},0}$ & $\pi_{\mathrm{B},-1}$
\end{tabular}
$
\ \par
\ \par
\ \par
Thus the logical qubit state "0", that is $\left|\Downarrow \Uparrow\uparrow\downarrow\right>$, is formed in
the following way (the pulses act on underlined spins from the left):
\begin{eqnarray}
\begin{tabular}{c c c c c}
$ \underuparrowA  \downarrowB  \uparrowA  \downarrowB \ldots \pi_{\mathrm{A},-1/2}$
& $\Rightarrow$ & $\DownarrowA  \underdownarrowB  \uparrowA  \downarrowB  \uparrow \ldots \pi_{\mathrm{B},0}$
& $\Rightarrow$ & $\mathrel{\mathop{\doubleoverline{\DownarrowA  \UparrowB  \uparrowA  \downarrowB }}\limits^{"0"}} \uparrow \ldots$ \nonumber
\end{tabular}
\end{eqnarray}
As the {\it ports}\/ (noted in Fig.\ 5 by cross `$\times$') for input and output
of the information in the array of ground states spin-qubits can be 
also dopant nuclei D at the certain place of the array with distinct 
resonant frequency, defects or local gates that modify the resonant 
frequency of the nearest nuclear spin in the array.\par
\par
\begin{center}
\epsfbox{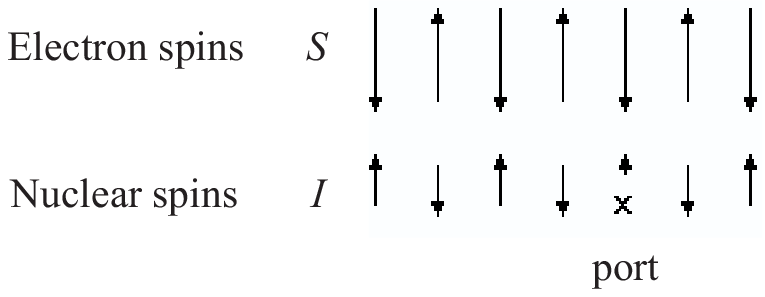}
\nobreak\par\nobreak
Fig.\ 5. Scheme of electron and nuclear spins ordering.\par
\end{center}
\par
Starting from the perfectly initialized states inputting the
information can be performed by setting the dopant D-spin to a desired
state by means of RF-pulse at its resonant frequency. The nuclear spin
state of spin nearest to the dopant spin is created by SWAP operation.
After the required information is loaded, D-spin is reset to the
ground state. Upon completion of computation, the state of any spin
can be measured by moving it to the A-site nearest to D, then swapping
$\mathrm{A}\Rightarrow \mathrm{D}$ and finally measuring the state of D-spin.\par
For the implementation of quantum operations on logic qubits we
will also introduce, one auxiliary {\it control unit}\/ (CU), which is
represented here by {\it six}\/ physical spin states in the pattern
$\doubleoverline{\Uparrow \Downarrow \downarrow  \uparrow  \Uparrow \Downarrow}$.
The CU exists only in one place along the array and is 
separated from logical qubits by odd number of spacer spins. The 
applying of corresponding SWAP sequence of pulses CU leads to putting 
the interaction of CU with one and two logical qubits and performing 
on them one- and two-qubit quantum operations (for more details, see 
\cite{13,25,26}).\par
In the case of a large enough ensemble of in parallel acting 
chains the states may be measured by NMR methods. For the increasing 
of logic qubit number in single `molecule' to $\geq 10^{3}$, that falls on one 
port, two- and three-dimensional structures with antiferromagnetic 
chess-type ordering of electron spins may be used. The corresponding 
ordering will also be for nuclear spins (Fig.\ 6).\par
\par
\begin{center}
\epsfbox{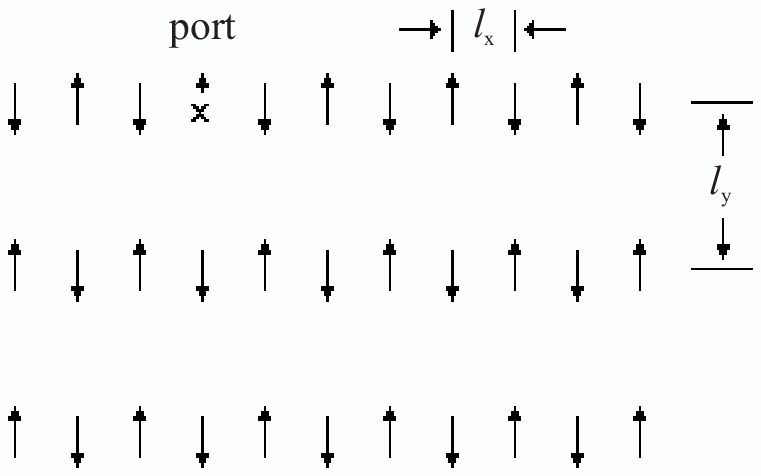}
\nobreak\par\nobreak
Fig.\ 6. Scheme of two dimensional chess-type ordering of initialized nuclear spins.\par
\end{center}
\par
If, for example, the number of spins-qubits, falls on one port in
linear chain is, say, $L \sim 30$, so in two-dimension case their number
will be $L = 900$. Ensemble that is composed of $N \sim 10^{5}$ in parallel
acting such plane `artificial molecules' permit to provide input and
output of information through the standard NMR techniques. The using
of more sensitive ENDOR techniques has particular meaning when it may
be combined with the techniques of dynamic polarization (solid state
effect).\par
Structures with two and three-dimensional antiferromagnetic order
may be found perhaps among the {\it natural rare earth}\/ or {\it transition
element}\/ dielectric compounds.\par
There are the rare earth compounds of {\it thulium}\/ stable isotope
$^{169}\mathrm{Tm}$, that has nuclear spin $I = 1/2$, $g_{\mathrm{N}} = 0.458$ and makes up 100\% of
abundance, with stable spinless isotopes of other elements Tm. They
can be possible: Tm$_{2}\mathrm{O}_{3}$, TmSi$_{2}$, TmGe$_{2}$, TmSe. The natural elements O,
Si, Ge and Se have, accordingly, nuclear spin containing isotopes (in
brackets the isotope abundance is shown) $^{17}\mathrm{O}$ $I = 5/2$ (0.04\%), $^{29}\mathrm{Si}$
$I = 1/2$ (4.7\%), $^{73}\mathrm{Ge}$ $I = 9/2$ (7.76\%), $^{77}\mathrm{Se}$ $I = 1/2$ (7.78\%). The
choosing the needed compounds requires further detailed theoretical
and experimental investigations.\par
It may be also considered, as variant for ensemble NMR quantum
computer, organic dielectric crystals, containing quasi-one-dimensional chains, such as antiferromagnetically ordered chains of
polyacetylene with only proton nuclei, for qubits:\par
\begin{center}
\epsfbox{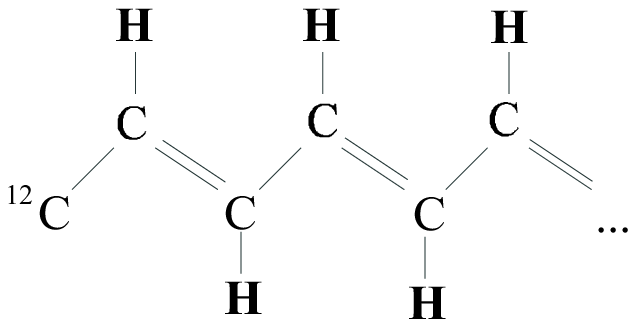}
\end{center}
\par
The advantages of ensemble quantum cellular automaton in
comparison with the above considered ensemble variant with strip gates
are as follows:\par
\begin{enumerate}
\item[a)] The system of the control gates is absent, what essentially
simplifies the production of computer structure and eliminates one of
the important sources of decoherence.\par
\item[b)] The coding of logic qubits into four physical qubits gives a
higher degree of fault-tolerance in logic operations.\par
\end{enumerate}
It follows that the going to ensemble quantum cellular automaton
permits to overcome {\it the third and fourth}\/ difficulties of Kane's 
scheme.\par
The chief disadvantage of cellular automaton scheme is the 
relative complexity of logic operation performance.\par

\section*{Conclusion}
\begin{enumerate}
\item The line of the large-scale ensemble NMR quantum computer
development has certain advantage over Kane's scheme. It consists in
the possibility of employment of the standard NMR technique for the
measurement of quantum states at output of computer, like in the
liquid prototype.\par
\item For the initialization of nuclear spin states at temperature
$T \sim 0.1\,\mathrm{K}$ methods of dynamic polarization may be proposed.\par
\item Analysis of proposed planar structure of ensemble silicon
computer shows the possibility of realization of large-scale NMR
quantum computer for ensemble component number $N \sim 10^{5}$.\par
\item The main reasons for the internal decoherence of one qubit
states are the modulation of resonance qubit frequency due to
hyperfine interaction with fluctuating electron spin and due to
interaction with randomly distributed impurity diamagnetic atoms
containing nuclear spins.\par
\item Analysis of different feasible ways for obtaining decoherence
times large enough shows that the values, needed to perform the
required for large-scaled computations number of quantum logic
operations $\sim 10^{5}$, can be achieved.\par
\item The implementation of cellular automaton principle permits to
abandon the realization regular nanostructure in the form of gate
chains.\par
\end{enumerate}

\section*{Appendix}
\par
\subsection*{A1. Signal NMR for discrete ensemble of nuclear spins}
\par
Let us consider here the sample that involves $N$ = $nL\cdot pN_{0}$ nuclear
spin--qubits arranged in the plane $z = 0$ of the silicon plate at
regular intervals along the strips (Fig.\ 4). The spins in chains under
strip at resonance in each block are oriented along $x$ axis (solenoid
axis) and separated at intervals of $L$. The read out NMR signal is
\begin{eqnarray}
\left|V_{\mathrm{max}}\right| = Q \omega_{\mathrm{A}} \frac{K}{X} \int_{-X/2}^{X/2} \left|\int_{A} B_{\mathrm{xma}\mathrm{x}}(x,y,z) dydz \right| dx ,\label{A.1}
\end{eqnarray}
where $X = nL \gg L$ is roughly the length of solenoid and
\begin{eqnarray}
B_{\mathrm{x}\; \mathrm{max}}(x,y,z) = \frac{\mu_{0}\gamma_{\mathrm{I}}\hbar }{16\pi } {\sum_{n_{\mathrm{i}}=-n/2}^{n/2}\;} {\sum_{p_{\mathrm{i}}=-pN_{0}/2}^{pN_{0}/2}} \frac{- 2(x - Ln_{\mathrm{i}})^{2} + (y - l_{\mathrm{y}}p_{\mathrm{i}})^{2} + z^{2}}{[(x - Ln_{\mathrm{i}})^{2} + (y - l_{\mathrm{y}}p_{\mathrm{i}})^{2} + z^{2}]^{5/2}}\label{A.3}
\end{eqnarray}
is the peak magnetic field produced by resonant spins in solenoid. For
simplicity it is suggested that $n, p, N_{0}, L \gg 1$ are the even numbers.
We assume that the area of coil turns is A = $D\cdot \delta $ ($D = l_{\mathrm{y}}\cdot pN_{0}$, $\delta \ll D$).\par
For summation over $n_{\mathrm{i}}$ and $p_{\mathrm{i}}$ we have used the {\it Poisson summation
formula}, namely,
\begin{eqnarray}
{\sum_{p_{\mathrm{i}}=-pN_{0}/2}^{pN_{0}/2}} f(l_{\mathrm{y}}p_{\mathrm{i}}) = \frac{pN_{0}}{D} {\sum_{\nu =-\infty }^{\infty }} \int_{-D/2}^{D/2} f(\xi ) \exp (i\nu 2\pi \xi /l_{\mathrm{y}}) d\xi ,\label{A.4}
\end{eqnarray}
by omitting the oscillated terms with $\nu \neq 0$:
\begin{eqnarray}
\left|V_{\mathrm{max}}\right| = \mu_{0} QK \omega_{\mathrm{A}} \frac{\gamma_{\mathrm{I}}\hbar }{8\pi } \frac{1}{X} \int_{-X/2}^{X/2}dx \frac{n}{X} \int_{-X/2}^{X/2}d\eta \cdot \nonumber
\end{eqnarray}
\begin{eqnarray}
\cdot \int_{0}^{\delta /2} dz \left| \frac{pN_{0}}{D} \int_{-D/2}^{D/2} \int_{-D/2}^{D/2} \frac{- 2(x - \eta )^{2} + (y - \xi )^{2}+ z^{2}}{[(x - \eta )^{2}+ (y - \xi )^{2}+ z^{2}]^{5/2}} dyd\xi \right|\label{A.5}
\end{eqnarray}
Taking into account $D^{2} \gg z^{2}$, upon integrating (\ref{A.4}) over $y$, $\xi$,
we obtain
\begin{eqnarray}
\left|V_{\mathrm{max}}\right| & = & \mu_{0}QK \omega_{\mathrm{A}} \frac{\gamma_{\mathrm{I}}\hbar }{8\pi } \frac{pN_{0}}{D} \frac{1}{X} \int_{-X/2}^{X/2} dx \frac{n}{X} \int_{-X/2}^{X/2}d\eta \cdot \nonumber
\\
&&  \cdot \int_{0}^{\delta /2} dz \left( \frac{2[(x-\eta )^{2}-z^{2}] D^{2}}{[(x-\eta )^{2}+ z^{2}]^{2} [D^{2}+(x-\eta )^{2}]^{1/2}} + \frac{2 z^{2}}{[(x-\eta )^{2}+ \mathrm{z}^{2}]^{3/2}} \right).\label{A.6}
\end{eqnarray}
By integrating now over $z$ and preserving only the major
logarithmically increasing for 2$\left|x - \eta\right|/\delta \Rightarrow 0$ term, we obtain
\begin{eqnarray}
\left|V_{\mathrm{max}}\right| & \approx & \mu_{0}{\it QKA }\omega_{\mathrm{A}} \frac{\gamma_{\mathrm{I}}\hbar }{4\pi } \frac{npN_{0}}{AD} \frac{1}{X} \int_{-X/2}^{X/2} dx \frac{1}{X} \int_{-X/2}^{X/2} d\eta \log \frac{\delta }{\left|x - \eta\right|} =\nonumber
\\
& = & (\mu_{0}/4) {\it QKA }\omega_{\mathrm{A}} \frac{N}{V_{\mathrm{s}}} \gamma_{\mathrm{I}}\hbar \cdot \frac{X}{\pi D} \log \frac{X}{\delta \sqrt{\mathrm{e}}},\label{A.7}
\end{eqnarray}
where $V_{\mathrm{s}} = AX$, $N = npN_{0}$, e = 2.718\ldots.\par
We see that expression (A.7) is distinguished from (\ref{4}) by non-essential factor $\frac{X}{\pi D} \log \frac{X}{\delta \sqrt{\mathrm{e}}}$, that is of the order of several ones.\par
\par
\subsection*{A2. Semiclassical model of adiabatic decoherence of one-qubit state}
\par
We will consider a long-lived non-equilibrium qubit state when
diagonal elements of density matrix may be treated as a constant.\par
The random modulation of resonance frequency $\Delta \omega (t)$ that causes
the dephasing of a qubit state is determined by the random phase
shifts
\begin{eqnarray}
\varphi (t) = \int_{0}^{t}\Delta \omega (t)dt.\label{A.8}
\end{eqnarray}
The one-qubit density matrix of pure state in rotating frame with
non perturbed resonance circular frequency will be
\begin{eqnarray}
\rho (t) = 1/2\left[
\begin{tabular}{c c }
$1 + P_{\mathrm{z}}$ & $P_{-}\exp (i\varphi (t))$\\
$P_{+}\exp (-i\varphi (t))$ & $1 - P_{\mathrm{z}}$\\
\end{tabular}
\right],\label{A.9}
\end{eqnarray}
where $P_{\pm } = P_{\mathrm{x}} \pm iP_{\mathrm{y}}$, $P_{\mathrm{x}}, P_{\mathrm{y}}, P_{\mathrm{z}}$ are Bloch vector components of length
$P = \sqrt{P_{\mathrm{x}}^{2} + P_{\mathrm{y}}^{2} + P_{\mathrm{z}}^{2}} = 1$.\par
By treating the resonance frequency modulation as Gaussian random
process after averaging (\ref{A.9}) over phase distribution with $\left<\varphi (t)\right> = 0$
we obtain
\begin{eqnarray}
\left<\rho (t)\right> = 1/2\left[
\begin{tabular}{c c }
$1 + P_{\mathrm{z}}$ & $P_{-}\exp (-\Gamma (t))$\\
$P_{+}\exp (-\Gamma (t))$ & $1 - P_{\mathrm{z}}$\\
\end{tabular}
\right],\label{A.10}
\end{eqnarray}
where
\begin{eqnarray}
\Gamma (t) = 1/2\cdot \left<\left(\int_{0}^{t}\Delta \omega (t)\mathrm{d}t\right)^{2}\right> = \int_{0}^{t}(t-\tau) \left<\Delta \omega (\tau )\Delta \omega (0)\right>d\tau ,\label{A.11}
\end{eqnarray}
$f(t) = \left<\Delta \omega (t)\Delta \omega (0)\right>$ is the frequency correlation function of a random
process, which is characterized by variance $\left<\Delta \omega (0)^{2}\right>$ and correlation
time $\tau_{\mathrm{C}}$ such that for $t > \tau_{\mathrm{C}} \left<\Delta \omega (t)\Delta \omega (0)\right> \Rightarrow 0$. For $\Gamma (t) > 0$ the
averaged density matrix presents a mixed quantum state with two non-zero eigen states
\begin{eqnarray}
1/2\cdot \left( 1 \pm \sqrt{1 - (P_{\mathrm{x}}^{2} + P_{\mathrm{y}}^{2})(1-\exp (-2\Gamma (t))}\right)\label{A.12}
\end{eqnarray}
and the populations of states $p_{\pm } = 1/2(1 \pm P_{\mathrm{z}}(0))$ at $\Gamma (t) \Rightarrow \infty $.\par
Thus, the adiabatic decoherence problem is reduced to the 
determination of the function $\Gamma (t)$ or the correlation function of 
random frequency modulation.\par
In the case of an ensemble quantum register there is a need to
average the one-qubit density matrix and correlation function over
ensemble of independent equivalent spins-qubits.\par
\par

\end{document}